# Local probing of ionic diffusion by electrochemical strain microscopy: spatial resolution and signal formation mechanisms

A.N. Morozovska,[1,] E.A. Eliseev,[2] N. Balke[3], and S.V. Kalinin[3,*]

[1] Institute of Semiconductor Physics, National Academy of Sciences of Ukraine,
41, pr. Nauki, 03028 Kiev, Ukraine

[2] Institute for Problems of Materials Science, National Academy of Sciences of Ukraine,
3, Krjijanovskogo, 03142 Kiev, Ukraine

[3] The Center for Nanophase Materials Sciences, Oak Ridge National Laboratory,
Oak Ridge, TN 37922

Electrochemical insertion-deintercalation reactions are typically associated with significant change of molar volume of the host compound. This strong coupling between ionic currents and strains underpins image formation mechanisms in electrochemical strain microscopy (ESM), and allows exploring the tip-induced electrochemical processes locally. Here we analyze the signal formation mechanism in ESM, and develop the analytical description of operation in frequency and time domains. The ESM spectroscopic modes are compared to classical electrochemical methods including potentiostatic and galvanostatic intermittent titration (PITT and GITT), and electrochemical impedance spectroscopy (EIS). This analysis illustrates the feasibility of spatially resolved studies of Li-ion dynamics on the sub-10 nanometer level using electromechanical detection.

---

[*] sergei2@ornl.gov

## 1. Introduction

Energy storage systems are the key enabling component of hybrid and electric automotive systems, portable electronics, and renewable-energy based energy technologies.[1,2,3] Crucial for progress to understand, develop, and optimize battery materials is the capability to decipher individual mechanisms responsible for battery functionality, including Li-ion and electron transport and electrochemical kinetics *locally*, at the level of grain assemblies, sub-micron grains, and ultimately at the nanometer scale of individual structural and morphological defects. Significant progress in this direction has been achieved with optical and micro-Raman imaging of battery materials and *in-situ* operational devices. However, the spatial resolution of optical methods is generally limited to the ~0.3 – 1 micron level, precluding the studies of energy materials below single-grain level. The quest for high-resolution probing of battery functionality has motivated a number of Scanning Probe Microscopy (SPM) based studies.

The local-probe SPM methods are generally based on force- or current detection. Studies of energy storage materials and phenomena hence necessitates coupling of the SPM signal to the particular aspect of electrochemical functionality. In liquids, a broad spectrum of current-based electrochemical SPM techniques have emerged based on the concept of ultramicroelectrodes, including techniques such as electrochemical scanning tunneling microscopy and atomic force microscopy,[4,5] and scanning ion conductance microscopy.[6,7] These families of SPM methods provide detailed information on the atomic and mesoscopic structure and morphology of solid-liquid interfaces and kinetics and thermodynamics of interfacial processes and liquid layer adjacent to the interface.

However, characterization of ionic diffusion *within the solid* represented a far more challenging task. Depending on operation mode, the AFM has been used to ascertain (a) the evolution of surface morphology during the charge-discharge cycles, (b) probe local static strains during the electrochemical processes, (c) local mapping of dc conductive currents and (d) SPM-based impedance imaging.

**(a) Topographic AFM:** Cohen and Aurbach used topographic AFM imaging to identify possible origins of capacity fading in the systems Li/LiPF$_6$(PC)/V$_2$O$_5$.[8] During intercalation, nano-sized LiF particles appeared at the V$_2$O$_5$ grain boundaries which decrease the Li flux in the battery systems resulting in slowing down Li-ion kinetics. Similarly, Doi et al. also identified the formation of small particles on $LiMn_2O_4$ as possible origin for capacity fading at elevated temperatures.[9] Overall, the capability of AFM to resolve the minute details of surface structure is invaluable in observation electrochemical processes; however, the chemical identification is generally unavailable and information is limited to surfaces only (as opposed to the details of ion insertion and dynamics within the material).

**(b) Static strain mapping:** The AFM can be used to map the strains developed in material during the ion intercalation that manifest as the shape changes of intrinsic or fabricated topographic features. The Dahn group used the AFM to measure volume changes during voltage cycles by using patterned electrode structures. [10,11,12] Beaulieu et al. and Matsui et. al used the AFM to measure roughness evolution during charge-discharge cycle of non-patterned electrode films of Si-Sn[13] and $LiCoO_2$[14] where direct height changes cannot be determined. The elegant work of Shao Horn group has demonstrated direct measurement of step height on $LiCoO_2$ surface as a function of Li concentration.[15] This approach, however, is slow and is applicable only to a limited number of materials.

**(c) DC conductance mapping:** The SPM can be used as a moving current electrode, potentially extending well-known time domain electrochemical methods as PITT, GITT, and charge-discharge measurements, to the nanoscale. Recently, Semenov et al.[16] used a biased tip to image spatial distribution of conductance on $V_2O_5$ on top of $Li_3PO_4$ electrolyte. Kuriyama et al.[17] measured currents on a bare $LiMn_2O_4$ surface in air and measured locally current changes with a slowly increasing electrical field (0.05 V/min). The Li-ions in $LiMn_2O_4$ are extracted tip field from the tetrahedral sites, resulting in current increase and topographical changes ascribed to a relaxation of Jahn-Teller instability. The primary limitation of the SPM-based current detection is that the sum of ionic and electronic currents is measured. Given that standard (e.g. Pt or Au coated) SPM tip is blocking electrode and the fact that ionic impedance are typically very large compared to electronic, the information on ionic flows is essentially lost.

**(d) AFM based impedance measurements.** A number of authors[18,19,20] have demonstrated the use of an AFM as a probe for local electrochemical impedance spectroscopy. However, the simple comparison of the tip-surface junction and cantilever surface impedances illustrates that direct measurements are possible only for well-defined mesoscopic objects (i.e. single-crystalline conductive grain with insulating grain boundaries), but not local volume of material below the tip.

Overall, the existing strategies for SPM-based probing of electrochemical processes in solids do not allow high-veracity studies due to the limits in spatial resolution, lack of quantitativeness, and multiple mechanisms contributing to measured signal and hence spatially-resolved contrast. The use of ion-sensitive local electrodes (similar to liquid electrochemical SPM) to directly probe ionic currents in solids is limited by slow diffusion rates and large contact impedances that effectively limit spatial resolution to 10s of microns. Hence, the capability for probing electrochemical functionality in solids has been elusive.

Recently, we have proposed that local electrochemical dynamics in solids can be studied using bias-strain coupling mediated by ionic diffusion.[21,22,23] In this method, the periodically biased conductive SPM tip concentrates electric field in a small volume of material, resulting in redistribution of mobile ions through diffusion and electromigration mechanisms. The associated changes in molar volume and strains results in periodic surface displacement detected by an SPM tip. This approach is further referred to as electrochemical strain microscopy (ESM), and is similar to the well-known Piezoresponse Force Microscopy [24, 25, 26, 27] of ferroelectrics and multiferroics based on converse piezoelectric effect. Here, we analyze the image formation mechanism in ESM for the case of a single-step diffusion process, derive the local strain responses in frequency and time domains, and analyze the sensitivity and resolution limits.

## II. Principles and Image Formation mechanism of E-PFM

The fundamental principles of electrochemical PFM are illustrated in Fig. 1. The biased SPM tip concentrates the electric field within a small volume of material, resulting in Li (or other mobile ion) redistribution. Unlike ferroelectric and piezoelectric materials in which the local stresses are directly coupled to the local electric field, in the electrochemical materials strain distribution is controlled by the (non-local) ion dynamics. Depending on the details of the experimental set-up and voltage range used, the process can be induced with fully metallic (blocking electrode), in which case tip-induced electromigration (at low voltages) and lithium extraction and formation of metallic lithium (for high voltages) on the initial stages of imaging process are possible. In ambient, the process is mediated by the formation of liquid droplet on the tip-surface junction,[28,22] that serves as the lithium reservoir and renders the tip-electrode system reversible (e.g. for materials such as $LiCoO_2$). Alternatively, imaging can be performed in Li-containing electrolyte. Note that the use of high frequencies as discussed below effectively precludes stray electrochemical reactions even at high voltages, as recently demonstrated for liquid imaging of model ferroelectric systems.[29,30] Finally, the measurements can be performed in the standard thin-film battery configuration, with the bias applied between the cathode and anode and SPM tip detecting the periodic strain generated on the materials surface.

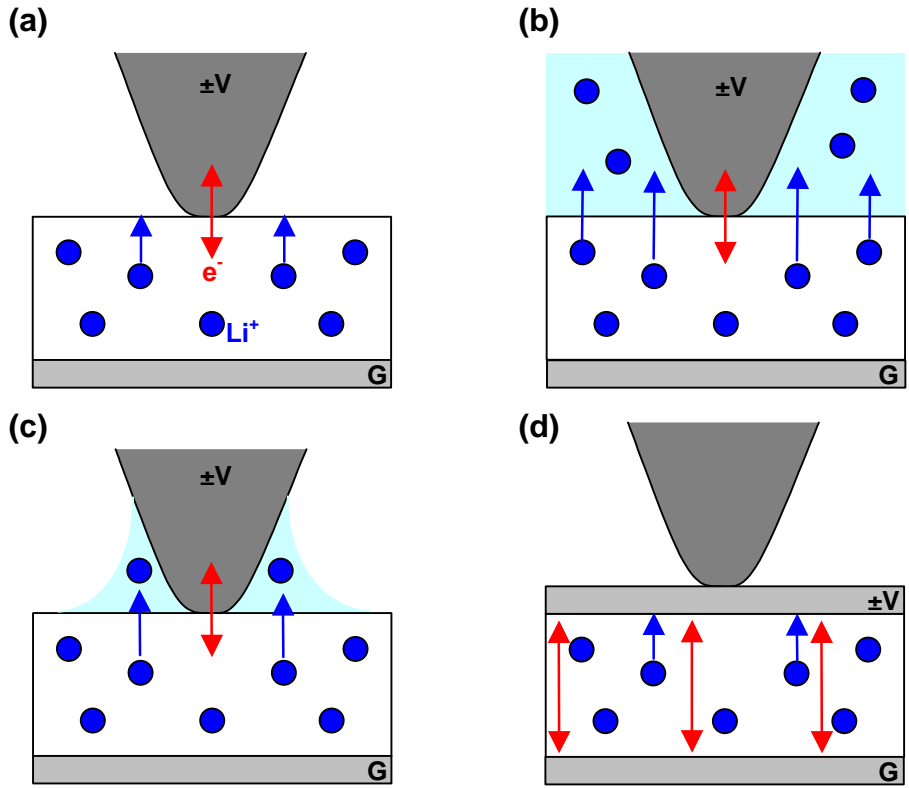

**Fig. 1.** Operational regimes for Electrochemical Strain Microscopy. (a) For blocking tip electrode, the electron transfer between the tip and the surface and non-uniform electrostatic field result in mobile ion redistribution within the solid, but no electrochemical process at the interface occurs. (b) In ambient conditions, the formation of the liquid droplet at the tip-surface junction provides a Li-ion reservoir, rendering electrode (partially) reversible. Similar effect can occur for blocking electrode at high biases (Li-extraction and tip plating) or for electrode coated by Li-electrolyte. (c) ESM can be performed in liquid Li-containing electrolyte (note that even for finite electronic conductivities ac electric field is concentrated in the tip-surface junction). (d) ESM can be performed on the surface of the top-electroded device. In cases a-b the electric field created by the probe is localized, in case d the field is uniform, in case c the field localization is controlled

by solution conductivity and modulation frequency. In all cases a-d, the tip detects local strain induced by local or non-local electric field.

Here, we analyze image formation mechanism for the case of fully reversible Li-ion electrode, corresponding to the case of Li-ion conductive material or process mediated by the liquid droplet in the tip-surface junction (i.e. Fig. 1 b or c for localized electric field). We adopt the method equivalent to the decoupled approximation previously used for ferroelectric materials.[31,32,33,34] In this case, (a) the lithium concentration is found ignoring the diffusion-strain coupling effects, (b) the local stresses are calculated using corresponding constitutive relations (Vegard law), and (c) strain and displacement fields in solid are calculated using appropriate Green's function. We further neglect inhomogeneous thermal expansion in comparison with chemical contribution. The mechanical displacement induced by Li-diffusion is studied in Section II.1. Electrostatic potential and constitutive equations are defined in Section II.2. Frequency dependences of the signal in ESM is analyzed in Sections III.1-2. Spatial resolution in ESM is discussed in Section III.3. Mechanical response in time domain is analyzed in Section IV. The detection and sensitivity limits are discussed in Section V. Note that while the examples are taken exclusively for Li-intercalation materials, similar derivations will be valid for other intercalation chemistries.

## II. 1. Mechanical displacement caused by ionic diffusion

The problem of mechanical stresses developing in the electrochemical systems have been recently addressed by a number of authors, including both the cases of macroscopic material and case of spherical particle [35, 36, 37, 38, 39]. For the latter, both decoupled [35] and coupled [37-39] numerical solutions are available. Importantly, the error induced by decoupling approximation is shown to be proportional to the squire of the molar expansion tensor and generally does not exceed 30% [38], well below the uncertainty of tip-surface contact radius in a typical SPM experiment.

Here, the mechanical stresses that develop during AFM-induced local electrochemical process were simulated for plane-stress conditions by solving the mechanical equilibrium equation, and mathematically describing the elastic contributions to the deformation of each volume element of material, as detailed by several authors [40, 41]. For the particular case when the chemical contribution is the dominant active mechanisms for strain, the equations of state (Hooke's law for the chemically active solid) for isotropic elastic media, subjected to the ionic flux relates concentration excess δ$C$, mechanical stress tensor $\sigma_{ij}$ and elastic strain $u_{ij}$ are the following [42, 43]:

$$u_{ij} = \beta_{ij}\, \delta C + s_{ijkl}\sigma_{kl}. \quad (1)$$

Here $s_{ijkl}$ is the tensor of elastic compliances, $\beta_{ij}$ is the Vegard tensor of chemical expansion, describing the lattice deformations under the small changes of composition δ$C$.

In subsequent analysis we note that the typical contact area in SPM experiment is well below micron-scale. The corresponding intrinsic resonance frequencies of material are thus in the GHz range, well above the practically important limits (both in terms of ion dynamic, and SPM-based detection of localized mechanical vibrations). Hence, we consider the general equation of mechanical equilibrium $\partial \sigma_{ij}/\partial x_j = 0$ in the quasi-static case that leads to the equation for mechanical displacement vector $u_i$ in the bulk of the system (see Appendix A) as:

$$c_{ijkl}\frac{\partial^2 u_k}{\partial x_j \partial x_l} - c_{ijkl} \cdot \beta_{kl}\frac{\partial \delta C}{\partial x_j} = 0. \quad (2)$$

Boundary conditions on the free surface $S$ are

$$\left(c_{ijkl}\frac{\partial u_k}{\partial x_l} - c_{ijkk}\beta_{kl}\, \delta C\right)n_j\bigg|_S = 0, \quad (3)$$

where $c_{ijkl}$ is the tensor of elastic stiffness, and $n_j$ is the components of the surface normal. General solution of the problem (2)-(3) is

$$u_i(x_1,x_2,x_3,t) = \iiint_{\xi_3>0}\frac{\partial G^S_{ij}(x_1-\xi_1,x_2-\xi_2,x_3,\xi_3)}{\partial \xi_m}c_{jmkl}\,\beta_{kl}\,\delta C(\xi_1,\xi_2,\xi_3,t)d\xi_1 d\xi_2 d\xi_3. \quad (4)$$

The essential condition for the existence of well-defined solution existence is the absence of δ$C$ at the infinity, well satisfied in an SPM experiment with local excitation (i.e. Fig. 1 a-c). $G^S_{ij}$ is appropriate tensorial Green function (see [44] and Appendix A). Here we approximate the symmetry of elastic properties as isotropic (well justified to 3D compounds such as spinels and olivines), albeit numerical schemes for Eq. (4) can be developed for lower symmetries in straightforward fashion. We further restrict the analysis to the transversally isotropic Vegard tensor $\beta_{ij} = \delta_{ij}\beta_{ii}$ with $\beta_{11} = \beta_{22} \neq \beta_{33}$ ($\delta_{ij}$ is the Kroneker delta symbol).

After elementary transformations of Eq. (4), the maximal surface displacement corresponding to the point $x_3=0$, i.e. surface displacement at the tip-surface junction detected by SPM electronics, for elastically isotropic semi-space is

$$u_3(x_1,x_2,0,t) = -\frac{1}{2\pi}\iiint_{\xi_3>0}\left(\beta_{11}\frac{\xi_3\left(2(1+\nu)\left((x_1-\xi_1)^2+(x_2-\xi_2)^2\right)-(1+2\nu)\xi_3^2\right)}{\left((x_1-\xi_1)^2+(x_2-\xi_2)^2+\xi_3^2\right)^{3/2}} + \frac{3\xi_3^3\beta_{33}}{\left((x_1-\xi_1)^2+(x_2-\xi_2)^2+\xi_3^2\right)^{5/2}}\right)\delta C(\xi_1,\xi_2,\xi_3,t)\,d\xi_1 d\xi_2 d\xi_3 \quad (5)$$

Here $\nu$ is the Poisson coefficient.

After Fourier transformation and using Percival theorem Eq.(5) is rewritten as

$$u_3(x_1,x_2,0,t) = -\int_{-\infty}^{\infty}dk_2\int_{-\infty}^{\infty}dk_1\int_0^{\infty}d\xi_3\left(\begin{array}{c}\exp(-ik_1x_1-ik_2x_2-k\xi_3)\times\delta\tilde{C}(k_1,k_2,\xi_3,t)\\ \times\left(\frac{\beta_{33}}{2\pi}(1+k\xi_3)+\frac{\beta_{11}}{2\pi}(1+2\nu-k\xi_3)\right)\end{array}\right) \quad (6a)$$

Here $k^2 = k_1^2 + k_2^2$, $\delta\tilde{C}(k_1,k_2,\xi_3,t)$ is the 2D Fourier image of the concentration field $\delta C(x_1,x_2,\xi_3,t)$. In the general case,

$$\tilde{u}_i(k_1,k_2,x_3) = \int_0^{\infty}d\xi_3\beta_{kl}c_{jmkl}\tilde{G}^s_{ij,m}(k_1,k_2,x_3,\xi_3)\delta\tilde{C}(k_1,k_2,\xi_3,t). \quad (6b)$$

For the isotropic Vegard tensor $\beta_{11} = \beta_{22} = \beta_{33} = \beta$ Eqs.(5)-(6) reduces to:

$$u_3(x_1,x_2,0,t) = -\frac{(1+\nu)}{\pi}\iiint_{\xi_3>0}\frac{\xi_3\,\beta\,\delta C(\xi_1,\xi_2,\xi_3,t)}{\left((x_1-\xi_1)^2+(x_2-\xi_2)^2+\xi_3^2\right)^{3/2}}d\xi_1 d\xi_2 d\xi_3$$
$$= -\frac{1+\nu}{\pi}\beta\int_{-\infty}^{\infty}dk_2\int_{-\infty}^{\infty}dk_1\int_0^{\infty}d\xi_3\exp(-ik_1x_1-ik_2x_2-k\xi_3)\delta\tilde{C}(k_1,k_2,\xi_3,t) \quad (7)$$

Thus, Eqs. (5) and (7) define the surface displacement at location (0,0) induced by the redistribution of mobile ion defined by $\delta C(\xi_1,\xi_2,\xi_3,t)$ field.

### II.2. Constitutive equations with boundary conditions

Electrostatic potential $V_e(\mathbf{x},t)$ distribution can be found self-consistently from the electrostatic Laplace's equations with the boundary conditions $V_e(x_1,x_2,x_3=0,t)=V_0(x_1,x_2,t)$ at the tip electrode $x_3 = 0$ and potential vanishing at infinity or remote bottom electrode. When the current density at the tip electrode $x_3 = 0$ appeared only due to the Li cations, a local deviation from equilibrium of the surface electrostatic potential, i.e. the overpotential, constitutes the driving force for the reaction to take place. For lithium such reaction is given by equation $Li^+ + e^- \leftrightarrow Li$. The rate of the ions transfer from the electrolyte to the active material phase is controlled by the Butler-Volmer relation [45, 46]. The transport kinetics model has been analyzed using a phase-field formulation that resulted in a set of coupled Cahn-Hilliard equations [45]. Theoretical description proposed by García *et al.* [47, 48] is based on the free-energy density formulation and includes variational principles.

Generally, Li-ion dynamics can be reduced to the ambipolar diffusion equation [49]:

$$\frac{\partial}{\partial t}\delta C(\mathbf{x},t) = D\left(\frac{\partial^2}{\partial x_1^2} + \frac{\partial^2}{\partial x_2^2} + \frac{\partial^2}{\partial x_3^2}\right)\delta C(\mathbf{x},t). \tag{8}$$

Diffusion coefficient is $D$. The diffusion constant may depend on the ionic concentration. Here, we ignore this dependence in order to develop analytical description of the process. Furthermore, this approximation is rigorous in the high-frequency regime of ESM, when the changes of ionic concentrations are minute.

Note that electrostatic potential does not contribute to the equation in the case of ambipolar diffusion (see [49] and Appendix B). In decoupling approximation the concentration is calculated ignoring the strain effects, i.e. the term proportional $div(C\mathrm{grad}(\beta_{ij}\sigma_{ij}))$ in Eq.(8), since their contribution lead into (4) is proportional to $\beta_{ij}^2$, while we consider only the terms linear on Vegard tensor $\beta_{ij}$.

Boundary conditions to Eq.(8) are the absence of the time-dependent part $\delta C(\mathbf{x},t)$ at infinity and the most general third kind boundary conditions including electric current in the contact area [50]:

$$\lambda \frac{\partial}{\partial x_3}\delta C(x_1, x_2, 0, t) - \eta \delta C(x_1, x_2, 0, t) = -V_0(x_1, x_2, t),$$

$$\delta C(x_1, x_2, x_3 \to \infty, t) \to 0, \quad \delta C(\mathbf{x},0) = 0, \quad C_c(\mathbf{x},0) = C_c^0. \tag{9}$$

Here $V_0(x_1, x_2, t)$ is the electrostatic potential distribution at the tip electrode $x_3 = 0$. This boundary conditions reduces to the case of either fixed concentration or fixed ionic flux at phenomenological exchange coefficient $\lambda = 0$ or $\eta = 0$, correspondingly.

The phenomenological exchange coefficients $\lambda$ and $\eta$ can be expressed in terms of the materials constants (see Appendix B for details). In particular $\eta = FR_c S/Z_c$ originated from the ohmic term $U_0 = I_0 R_c = \delta C_c F \cdot S$ in the overpotential $U_e(x_1, x_2, t)$, while $\lambda \approx -eD \dfrac{RT(C_S - C_c^0)^{-\alpha_a}(C_c^0)^{-\alpha_c}}{F^2 \chi(\alpha_a + \alpha_c)}$ is determined by the reaction rate constant $\chi$ and the solubility limit of lithium in the tip electrode $C_S$. Here, $F$ is Faraday's constant, $R$ is the universal gas constant, $T$ is the absolute temperature, $\alpha_a$ is anodic empirical constant, $\alpha_c$ is cathodic empirical constant, $R_c$ is the total contact ohmic resistance, is the current, $\chi$ is the reaction rate, $C_S$ is the solubility limit of lithium in the electrode, $S$ is the cross-section area.

Using Laplace transformation on time $t$, and Fourier transformation on transverse coordinates, the solution of problem (12) was found as

$$\delta \tilde{C}(k_1, k_2, x_3, t) = \frac{1}{2i\pi} \int_{A-i\infty}^{A+i\infty} ds \exp\left(-x_3 \sqrt{k^2 + s/D} + st\right) \frac{\tilde{V}_0(k_1, k_2, s)}{\lambda \sqrt{k^2 + s/D} + \eta}. \tag{10}$$

Here vector $\mathbf{k} = \{k_1, k_2\}$, its absolute value $k = \sqrt{k_1^2 + k_2^2}$; $\tilde{V}_0(\mathbf{k}, s)$ is the Fourier-Laplace image of $V_0(x_1, x_2, t)$.

### III. Frequency dependence of the signal in Electrochemical PFM
### III.1. Mechanical response frequency spectrum

In particular case of periodic bias variation $V_0(x_1, x_2, t) \sim V_0(x_1, x_2) \exp(i\omega t)$ with temporal frequency $\omega$ and spectrum $\tilde{V}_0(k_1, k_2, \omega)$, the concentration spectrum can be derived as

$$\delta \tilde{C}(k_1, k_2, x_3, \omega) = \exp\left(-x_3 \sqrt{k^2 + i\omega/D}\right) \frac{\tilde{V}_0(k_1, k_2, \omega)}{\lambda \sqrt{k^2 + i\omega/D} + \eta}. \tag{11}$$

Here the condition $\text{Re}\left(\sqrt{k^2 + i\omega/D}\right) > 0$ should be valid to ensure the stability.

To define the boundary conditions in a generalized ESM experiment, we assume that the potential (and thus ionic flux) spatial distribution $V_0(x_1, x_2, t)$ is kept constant inside the circle of radius $R_0$ and zero outside. This condition provides an approximate description of the probe tip having a well-defined characteristic size. We further utilize the fact approximate solutions developed here are insensitive to the details of the probe shape. Hence, Fourier image can be taken as $\tilde{V}_0(k_1, k_2, \omega) = V(\omega) R_0 J_1(kR_0)/k$.

Hereinafter (unless specifically indicated) we consider the simplest case of the isotropic Vegard tensor $\beta_{11} = \beta_{22} = \beta_{33} = \beta$. Then allowing for Eq.(11), Eq.(7) for the surface mechanical displacement acquires the form

$$u_3(r,\omega) = \int_0^\infty dk \frac{-2(1+\nu)\beta V(\omega) \cdot R_0 J_1(kR_0) J_0(kr)}{\left(k + \sqrt{k^2 + i\omega/D}\right)\left(\lambda\sqrt{k^2 + i\omega/D} + \eta\right)}, \tag{12}$$

where the polar radius $r = \sqrt{x^2 + y^2}$. Below, we explore the specific limiting cases of Eq. (12) that define frequency response and losses in the ESM signal $u_3(0,\omega)$ and radial dependence of surface deformation profile, $u_3(r,\omega)$, i.e. spatial resolution of ESM.

### III.2. Electrochemical strain microscopy signal

The frequency-dependent strain signal in ESM provides an analog of classical current-based electrochemical impedance spectroscopy. Here, we analyze the frequency response of the signal, i.e. local electromechanical analog of diffusion Wartburg impedance. The important limiting cases for Eq. (12) is the case of $r = 0$, i.e. frequency dependent electromechanical response in the ESM. Analytical expressions for the response $u_3(0,\omega)$ are listed in the Table 1, which specifically indicates the limiting cases for low frequency and high frequency for flux-controlled and concentration-driven processes. Hereinafter the characteristic diffusion time $\tau = R_0^2/2D$ is introduced.

**Table 1. Frequency regimes for the Electrochemical Strain Microscopy response**

| | ESM response $u_3(0,\omega)$ for concentration-driven process ($\lambda = 0$) | ESM response $u_3(0,\omega)$ for flux-controlled process ($\eta = 0$) |
|---|---|---|
| Exact expression | $\dfrac{-(1+\nu)\beta V(\omega)}{\eta} \cdot \dfrac{R_0}{i\omega\tau}\left(\begin{array}{c}\exp\left(-(1+i)\sqrt{\omega\tau}\right)\\ -1+(1+i)\sqrt{\omega\tau}\end{array}\right)$ | $-\dfrac{(1+\nu)\beta V(\omega)}{\lambda} \dfrac{R_0^2}{i\omega\tau}\left(1 - G_{13}^{21}\left(\left.\dfrac{i\omega\tau}{2}\right\vert \begin{array}{c}0\\ -1/2, 1/2, -1/2\end{array}\right)\right)$ |
| $\omega\tau \ll 1$ | $-(1+\nu)\beta\dfrac{V(\omega)}{\eta}\left(1 - \dfrac{1+i}{3}\sqrt{\omega\tau}\right)R_0$ | $\dfrac{-(1+\nu)\beta V(\omega) R_0^2}{4\lambda}\ln\left(\dfrac{1}{2\omega\tau}\right)$ |

| $\omega\tau = 1$ | $\dfrac{-(1+\nu)\beta V(\omega)R_0}{\eta}\cdot(1.31-0.20i)$ | $-\dfrac{(1+\nu)\beta V(\omega)R_0^2}{\lambda}(0.30-0.28i)$ |
| $\omega\tau \gg 1$ | $(1+\nu)\beta\dfrac{V(\omega)}{\eta}\bigl(i+(1-i)\sqrt{\omega\tau}\bigr)\cdot\dfrac{R_0}{\omega\tau}$ | $\dfrac{-(1+\nu)\beta V(\omega)R_0^2}{\lambda}\dfrac{1}{i\omega\tau}\left(1-\sqrt{\dfrac{1}{i\omega\tau}}\right)$ |

Here $G_{13}^{21}\!\left(x\left|\begin{array}{c}0\\-1/2,1/2,-1/2\end{array}\right.\right)$ is Meijer G function.

The comparison of the exact and approximate expressions from the Table 1 is shown in Fig. 2. Here we introduced the dimensionless frequency $w=\omega R_0^2/D = 2\omega\tau$. It is seen that the approximate expressions describe the exact frequency dependence with sufficient accuracy in their applicability limits.

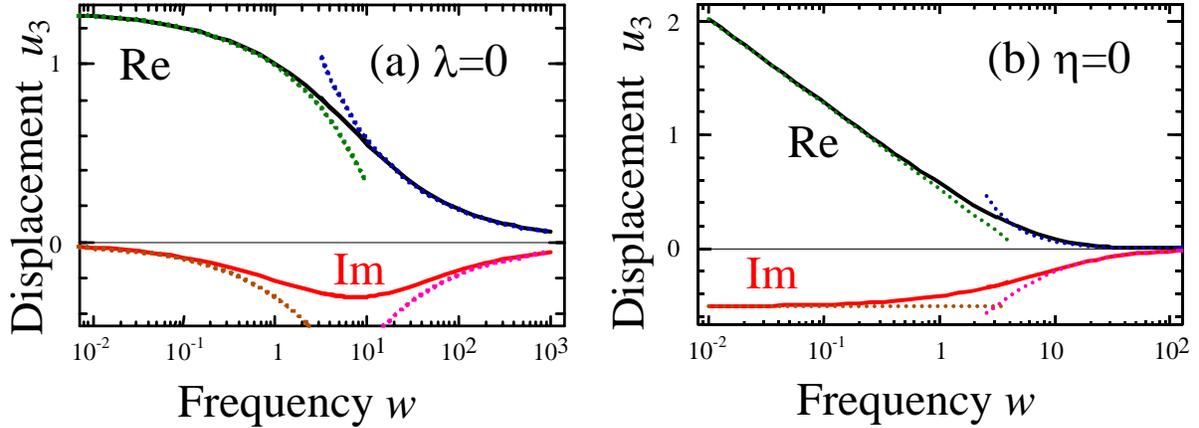

**Fig. 2.** The frequency dependence of electrochemical strain response for two limiting cases of boundary conditions, namely for (a) concentration controlled and (b) flux controlled cases. Solid and dotted lines represent exact and approximate expressions, respectively.

Using Eq.(11), the Fourier image of the concentration flux was derived as:

$$\left.\frac{\partial\delta\tilde{C}(k_1,k_2,x_3,\omega)}{\partial x_3}\right|_{x3=0} = -\frac{\sqrt{k^2+i\omega/D}}{\lambda\sqrt{k^2+i\omega/D}+\eta}\tilde{V}_0(k_1,k_2,\omega). \qquad (13)$$

Substituting the tip potential Fourier image $\tilde{V}_0(k_1,k_2,\omega)=V(\omega)R_0\,J_1(kR_0)/k$, the flux distribution at the surface $x_3 = 0$ and in the point **r** = 0 acquires the form

$$\left.\frac{\partial \delta C(r, x_3, \omega)}{\partial x_3}\right|_{x3=0} = -V(\omega)\int_0^\infty dk \frac{\sqrt{k^2 + i\omega/D} \cdot R_0 J_1(kR_0) J_0(kr)}{\left(\lambda\sqrt{k^2 + i\omega/D} + \eta\right)}. \tag{14}$$

$$\left.\frac{\partial \delta C(0, x_3, \omega)}{\partial x_3}\right|_{x3=0} \approx \frac{-V(\omega)\sqrt{i\omega/D}}{\lambda\sqrt{i\omega/D} + \eta\left(1 - \exp\left(-R_0\sqrt{i\omega/D}\right)\right)} \tag{15}$$

The latter approximation is exact in the limit of flux-controlled process, $\eta \to 0$.

The comparison of exact and approximate expressions for flux in the point $\mathbf{r} = 0$ is shown in Fig. 3.

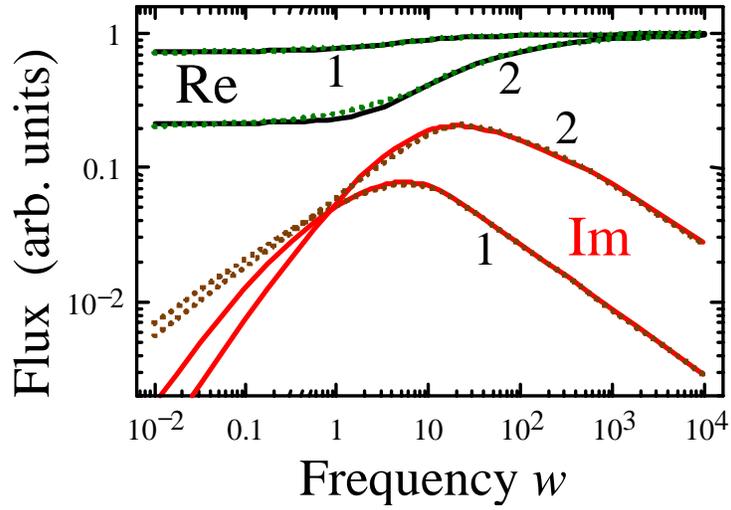

**Fig. 3.** The frequency dependence of the flux in the point $\mathbf{r} = 0$ (real and imaginary parts are shown) for $\eta = 0.4\lambda/R_0$ (curves 1) and $\eta = 4\lambda/R_0$ (curves 2). Solid and dotted curves represent exact and approximate expressions respectively.

The frequency dependence of the normalized ESM response, $\tilde{u}_3 = -u_3(0,w)(\eta + \lambda/R_0)/((1+\nu)\beta V(w)R_0)$, for the concentration- and flux-controlled processes is shown in Fig. 4.

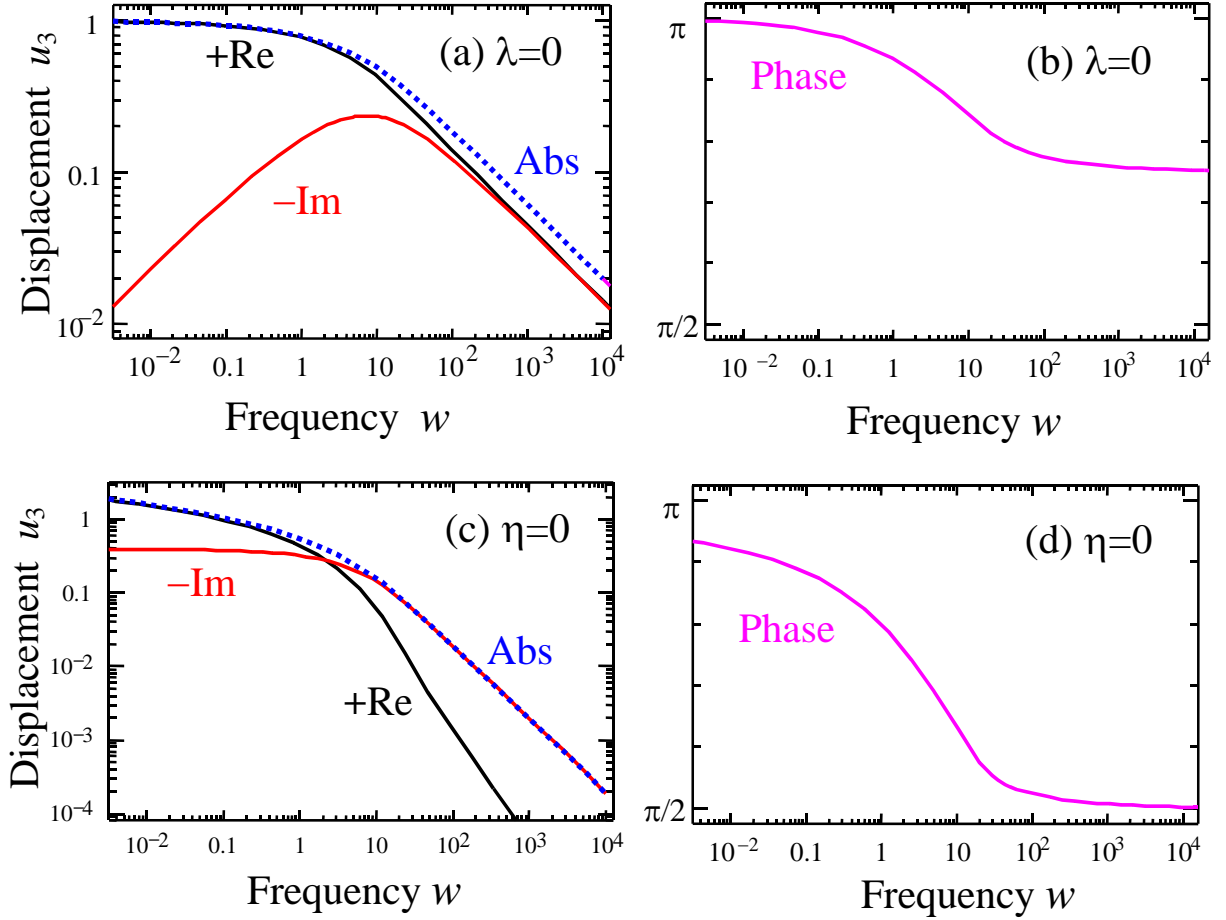

**Fig. 4.** The real (Re), imaginary (-Im) parts, absolute value (Abs) and phase of normalized displacement, $\tilde{u}_3$, vs. dimensionless frequency $w$ for determined concentration $\delta C$ at $\lambda=0$ (a,b) and fixed ionic flux $\partial \delta C / \partial x_3$ at $\eta=0$ (c,d).

When generating all subsequent plots we used exact integral in Eq. (12). In the case of concentration-driven process, the response is constant for small frequencies ($\omega\tau \ll 1$), and is linear in effective concentration $V(\omega)/\eta$ on the boundary. This independence of response on frequency (despite the fact that the diffusion length diverges for low frequencies) is related to the finite signal generation volume in ESM (i.e. strain contribution from the parts of material far from the tip-surface junctions is small). In comparison, the total flux is $\partial \delta \tilde{C}(k=0, x_3, \omega)/\partial x_3 \big|_{x3=0} = -\sqrt{i\omega/D}\, \tilde{V}_0(k=0,\omega)/\eta$. For high frequencies ($\omega\tau \gg 1$), the response decays as $(\omega\tau)^{-1}$.

In comparison, flux-controlled process ($\eta = 0$), the response logarithmically diverges for low frequencies, since the amount of transferred material increases linearly with cycle time. However, as a consequence of strain transfer effect this divergence is only logarithmic, as opposed to a power-law expected for current detection. For high frequencies, the response decays as $(\omega\tau)^{-1}$, and the corresponding phase angle becomes $\pi/2$.

It is instructive to analyze the dynamic response in terms of the hysteretic loop behavior, providing direct link to the observables in the SPM experiment. The complex quantities $\{\tilde{u}_3(\omega,t) = u_3(0,\omega)\exp(i\omega t), \ \tilde{V}_0(\omega,t) = V\exp(i\omega t)\}$ temporal behavior describes elliptic loop in the complex plane. The parametric dependence of observable quantity $\text{Re}[u_3(\omega)\exp(i\omega t)]$ on $\text{Re}[V(\omega)\exp(i\omega t)]$ describes elliptic loop at fixed frequency $\omega$ (see Figs. 5-6).

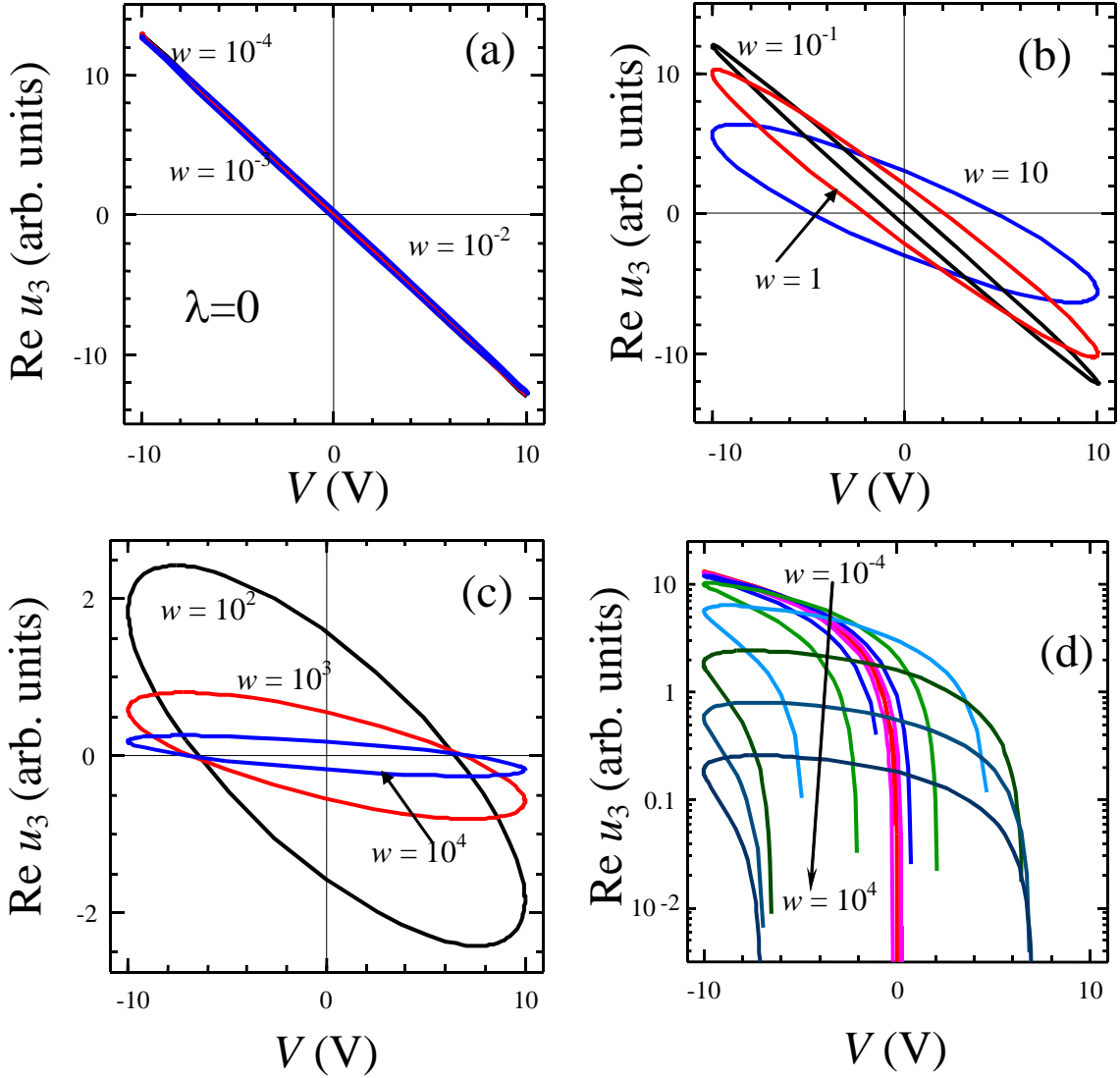

**Fig. 5.** Hysteresis loops for different dimensionless frequency $w$ (labels near the loops), $D = 10^{-12}$ SI units, $R_0 = 100$ nm and $\lambda=0$. (d) Semi-logarithmic plot of the loops in whole frequency interval (frequency increases with multiplier "10" from top to bottom).

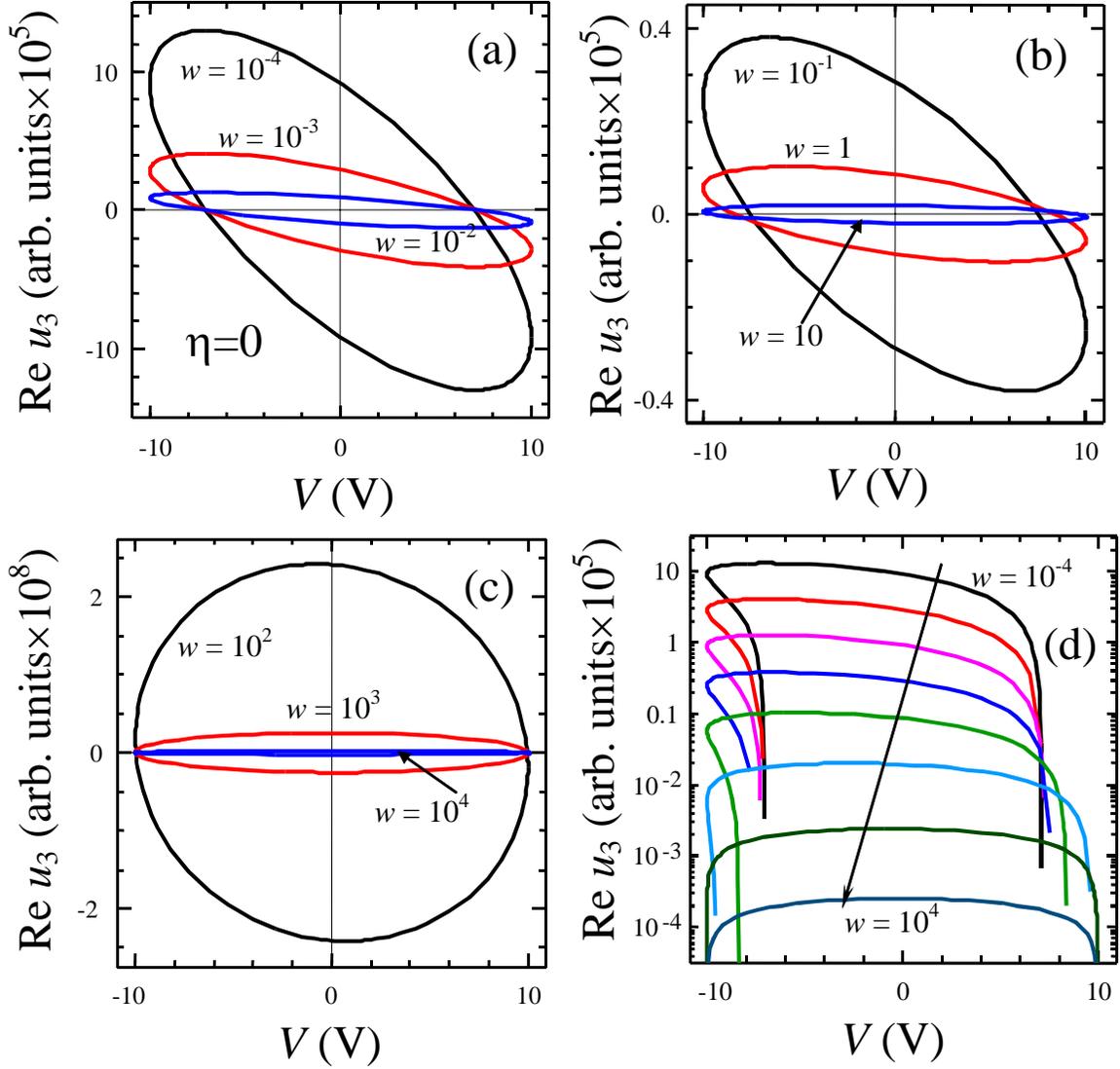

**Fig. 6.** Loops for different dimensionless frequency $w$ (labels near the loops), $D = 10^{-12}$ SI units, $R_0 = 100$ nm and $\eta=0$. (d) Semi-logarithmic plot of the loops in whole frequency interval (frequency increases with multiplier "10" from top to bottom).

As expected, the concentration driven process the hysteresis loop is closed and linear for small frequencies for which the ion-dynamics in the signal generation volume of material (of the order of $R_0$) follows probe bias exactly. For intermediate frequencies, $\omega\tau \approx 1$, the response is still large, but the ionic response becomes delayed compared to probe bias, resulting in hysteresis loop opening. Finally, for high frequencies strain response is small, and hence the hysteresis loop becomes effectively closed as well. In the flux-limited case, the hysteresis loop area diverges in the low-frequency limit, while in the high-frequency limit it is qualitatively similar to $\lambda=0$ case.

Analytically, the loop area $S$ is given by expression

$$S = \pi \cdot |V \cdot \mathrm{Im}(u_3(0,\omega))| \tag{16}$$

Loops area frequency dependence is shown in Fig. 7. As expected, the hysteresis loop area is maximum for concentration-driven case is maximal for $\omega\tau \approx 1$.

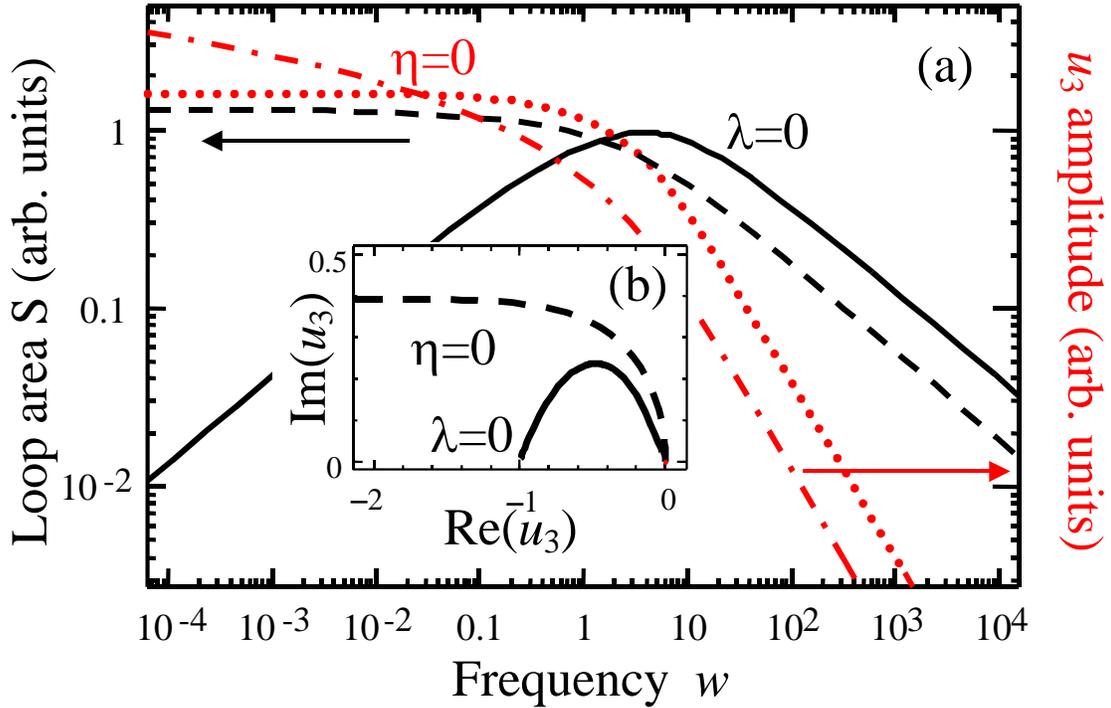

**Fig. 7.** (a) Loop area and absolute value of response vs. dimensionless frequency $w$ for $\eta=1$ and $\lambda=0$ (solid and dashed black curves), $\eta=0$ and $\lambda=1$ (dotted and dash-dotted red curves respectively). (b) Dependence of $\mathrm{Im}(u)$ vs. $\mathrm{Re}(u)$ for $\lambda=0$ (solid curve) and $\eta=0$ (dotted curve).

Note that this analysis is strictly valid only for small concentration changes, i.e. within the applicability limit of the linear diffusion equations. Experimentally, the response is expected to saturate once the concentration change becomes comparable to the total capacity of the material. Some of these effects are considered qualitatively in Section V.

### III.3. Spatial resolution in ESM

The spatial resolution in ESM can be estimated from the characteristic size of the area affected by the concentration change, i.e. lateral displacement profile on the surface. Simplified analytical expressions for the response radial dependence $u_3(r,\omega)$ are listed in the Table 2, which specifically indicates the limiting cases for low frequency and high frequency for flux-controlled and concentration-driven processes.

**Table 2. Spatial resolution in ESM measurements**

| | ESM response $u_3(r,\omega)$ for concentration-driven process ($\lambda = 0$) | ESM response $u_3(r,\omega)$ for flux-controlled process ($\eta = 0$) |
|---|---|---|
| $\omega\tau \ll 1$ | $\dfrac{-(1+\nu)\beta V(\omega)}{\eta}\dfrac{4R_0\left(1+(r/R_0)^3\right)}{4+(r/R_0)^2+8(r/R_0)^4}$ | $\dfrac{-(1+\nu)\beta V(\omega)}{\lambda(1+i)\sqrt{\omega\tau}}\dfrac{4R_0^2\left(1+(r/R_0)^3\right)}{4+(r/R_0)^2+8(r/R_0)^4}$ |
| $\omega\tau \gg 1$, $r \gg R_0$ $\tau = \dfrac{R_0^2}{2D}$ | $\mathrm{Im}[u_3] \approx \dfrac{(1+\nu)\beta V(\omega)}{4\eta\cdot\omega\tau}\left(1+\dfrac{9R_0^2}{8r^2}\right)\dfrac{R_0^4}{r^3}$  $\mathrm{Re}[u_3] \sim \dfrac{-(1+\nu)\beta V(\omega)}{\eta\sqrt{\omega\tau}}R_0\theta(R_0-r)$ | $\mathrm{Im}[u_3] \sim -\dfrac{(1+\nu)\beta V(\omega)}{4\lambda\cdot\sqrt{\omega\tau}}\left(1+\dfrac{9R_0^2}{8r^2}\right)\dfrac{R_0^5}{r^3}$  $\mathrm{Re}[u_3] \sim \dfrac{-(1+\nu)\beta V(\omega)}{\lambda\omega\tau}R_0^2\theta(R_0-r)$ |

The radial dependence of the response is shown in Fig. 8 for different frequencies. For low frequencies the lateral strain is generated over large areas, and far from the tip-surface junction the response decays as $\sim R_0^2/2r$ for concentration driven process and $\sim R_0^3/2r$ for flux driven process. As expected, in both cases the decay length is given by the characteristic tip size, $R_0$. It follows from the figure the excitation area is determined by several $R_0$ and tends to $R_0$ at high frequencies $w = \omega R_0^2/D = 2\omega\tau \gg 1$. The excitation depth (i.e. depth resolution) is determined by the lateral scale and characteristic frequency, since the elastic Green function of the semi-infinite medium has no characteristic scale. Based on the general physical considerations, the penetration depth can be estimated as probe radius for low frequencies ($w \ll 1$) and diffusion length for high frequencies ($w \gg 1$).

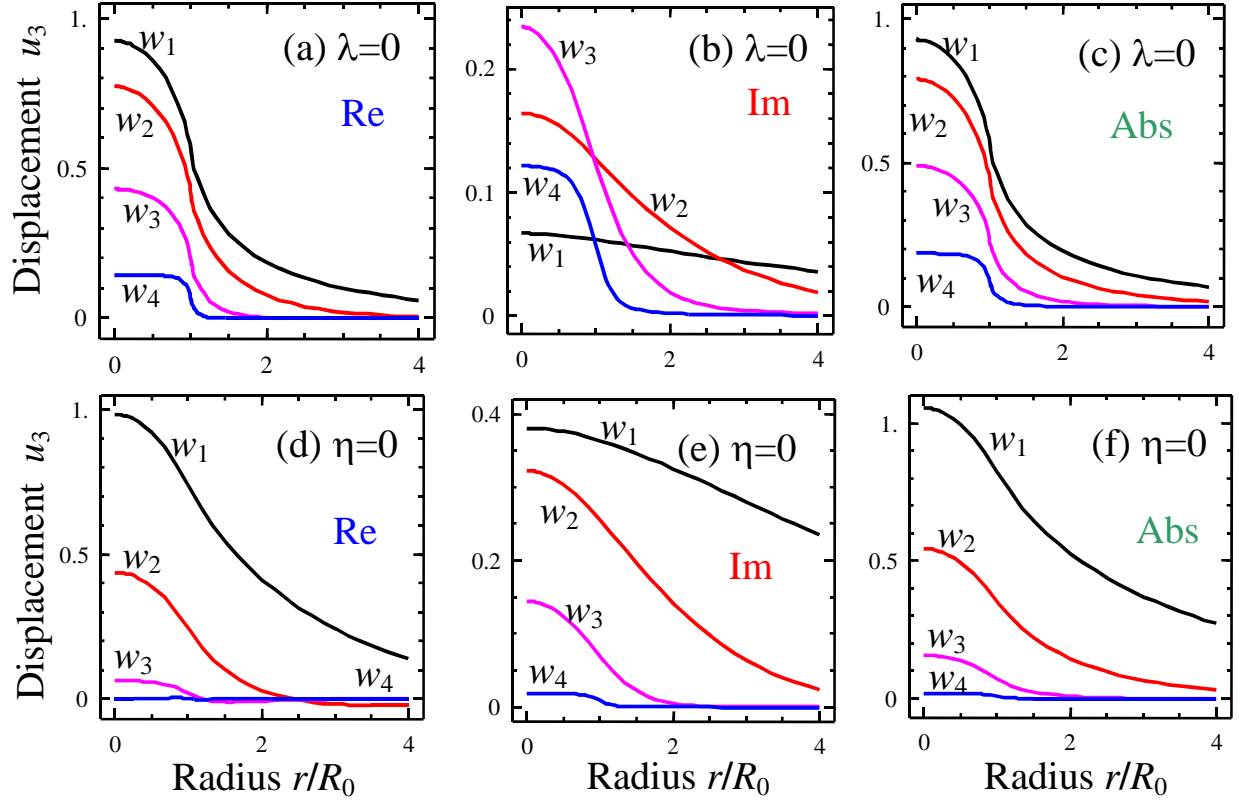

**Fig. 8.** Normalized displacement $u_3(r,w)$ real (Re) and imaginary (-Im) parts and absolute value (Abs) vs. dimensionless radius $r/R_0$ for different dimensionless frequency $w = 0.1, 1, 10, 100$ for fixed concentration $\delta C$ at $\lambda=0$ (a,b,c) and fixed ionic flux $\partial \delta C / \partial x_3$ at $\eta=0$ (d,e,f). When generating plots we used exact Eq.(12).

Here, we define the lateral resolution as the half width of the excited spatial region at half maximum (HWHM), normalized on tip size $R_0$. Fig. 9 shows the resolution as a function of frequency for limiting cases of boundary conditions with fixed concentration $\delta C$ ($\lambda=0$) and fixed ionic flux. The HWHM was determined as the half width of the absolute value (Abs) of response, thus it is different from the real (Re) and imaginary part (Im) half widths. As such, it was found to be dominated by the real part of the response (compare half widths in Figs.8a,b,d,e with the ones in Figs.8c,f).

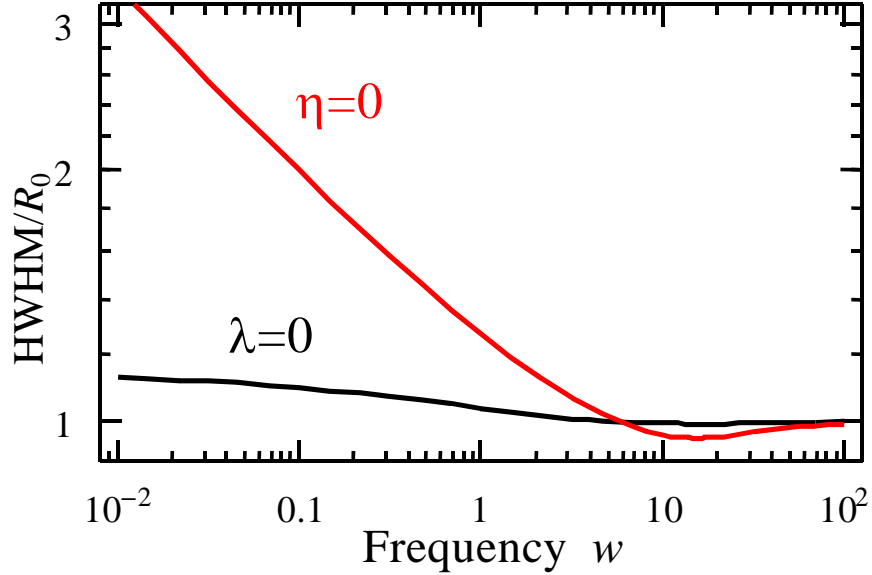

**Fig. 9** The frequency dependence of the excited spatial region halfwidth at half maximum (HWHM) normalized on tip size $R_0$ for two limiting cases of boundary conditions with fixed concentration $\delta C$ ($\lambda=0$) and fixed ionic flux $\partial \delta C/\partial x_3$ ($\eta=0$).

### IV. Mechanical response in time domain

The ESM measurements can be performed in the time-domain, in which the response is measured as a function of time during and after the application of the finite bias pulse. Dependent on the control variable (pulse length and amplitude, or final response), these measurements will be analogous to the potentiostatic- and galvanostatic intermittent titration (PITT and GITT), respectively.

Here, we analyze the time spectroscopy in the case of constant-potential time ESM spectroscopy. For the case of the potential spatial distribution $V_0(x_1, x_2, t) = const$ inside the circle of radius $R_0$ and zero outside, mechanical response in time domain can be derived from Eqs. (7) and (10) as

$$u_3(0,t) = -2(1+\nu)\beta \int_0^\infty dk \, \frac{1}{2i\pi^2} \int_{A-i\infty}^{A+i\infty} ds \, \frac{R_0 J_1(kR_0) \cdot \tilde{V}_0(s) \exp(+st)}{\left(\lambda\sqrt{k^2 + s/D} + \eta\right)\left(k + \sqrt{k^2 + s/D}\right)}. \qquad (17)$$

The analytical results were derived for the case of concentration-determined ($\lambda=0$) and flux-determined ($\eta=0$) process (see Appendix C). For the case $\lambda=0$ the relaxation of the surface displacement is:

$$u_3(0,t) = -2(1+\nu)\beta \int_0^t dt' \cdot V_0(t-t') \frac{D}{R_0 \eta}\left(\frac{R_0}{\sqrt{\pi D t'}} - \mathrm{erf}\left(\frac{R_0}{2\sqrt{Dt'}}\right)\right), \quad (18)$$

where $\mathrm{erf}(z) = \frac{2}{\sqrt{\pi}}\int_0^z \exp(-x^2)dx$ is the Gaussian error distribution.

For a rectangular voltage pulse, $V_0(t) = V_a(\theta(t) - \theta(t-t_0))$, with pulse duration $t_0$, approximate expressions were obtained from Eq.(20) as $u_3(0, t < t_0) \approx -(1+\nu)\frac{4\beta V_a}{\eta\sqrt{\pi}}\sqrt{Dt}$ and $u_3(0, t \gg t_0) \approx -(1+\nu)\frac{\beta V_a R_0^2 t_0}{6\eta\sqrt{\pi D}t^{3/2}}$. The response amplitude increases with $t_0$ as $u_3(0,t_0) \sim \sqrt{t_0}$ and decay when the pulse is off according to the power law $u_3(0, t \gg t_0) \sim t^{-3/2}$.

For the particular case $\eta=0$ the displacement signal obeys the law:

$$u_3(0,t) = -2(1+\nu)\beta \frac{D}{\lambda}\int_0^t dt' \cdot V_0(t-t')\left(1 - \exp\left(-\frac{R_0^2}{8Dt'}\right)I_1\left(\frac{R_0^2}{8Dt'}\right)\right), \quad (19)$$

where $I_1$ is the modified Bessel function.

For a rectangular voltage pulse, $V_0(t) = V_a(\theta(t) - \theta(t-t_0))$, with pulse duration $t_0$, the response increase with the bias pulse duration increase is logarithmic $u_3(0,t) \approx -(1+\nu)\beta V_a \frac{R_0^2}{4\lambda}\ln\left(1 + \frac{8Dt}{R_0^2}\right)$ at $t < t_0$, while the response decay when the bias pulse is turned off is quasi-logarithmic $u_3(0,t) \approx -(1+\nu)\beta V_a \frac{R_0^2}{4\lambda}\ln\left(\frac{R_0^2 + 8Dt}{R_0^2 + 8D(t-t_0)}\right)$ at $t > t_0$, and then tends to the power law $u_3(0, t \gg t_0) \sim t^{-1}$.

Displacement $u_3(0,t)$ relaxation in response to rectangular pulse for the cases of flux-controlled ($\eta=0$) and concentration-controlled ($\lambda=0$) process is shown in Fig. 10. Mechanical displacement increase and relaxation for the case $\eta=0$ appeared more sluggish than for the case $\lambda=0$ (compare plots c,d with a,b).

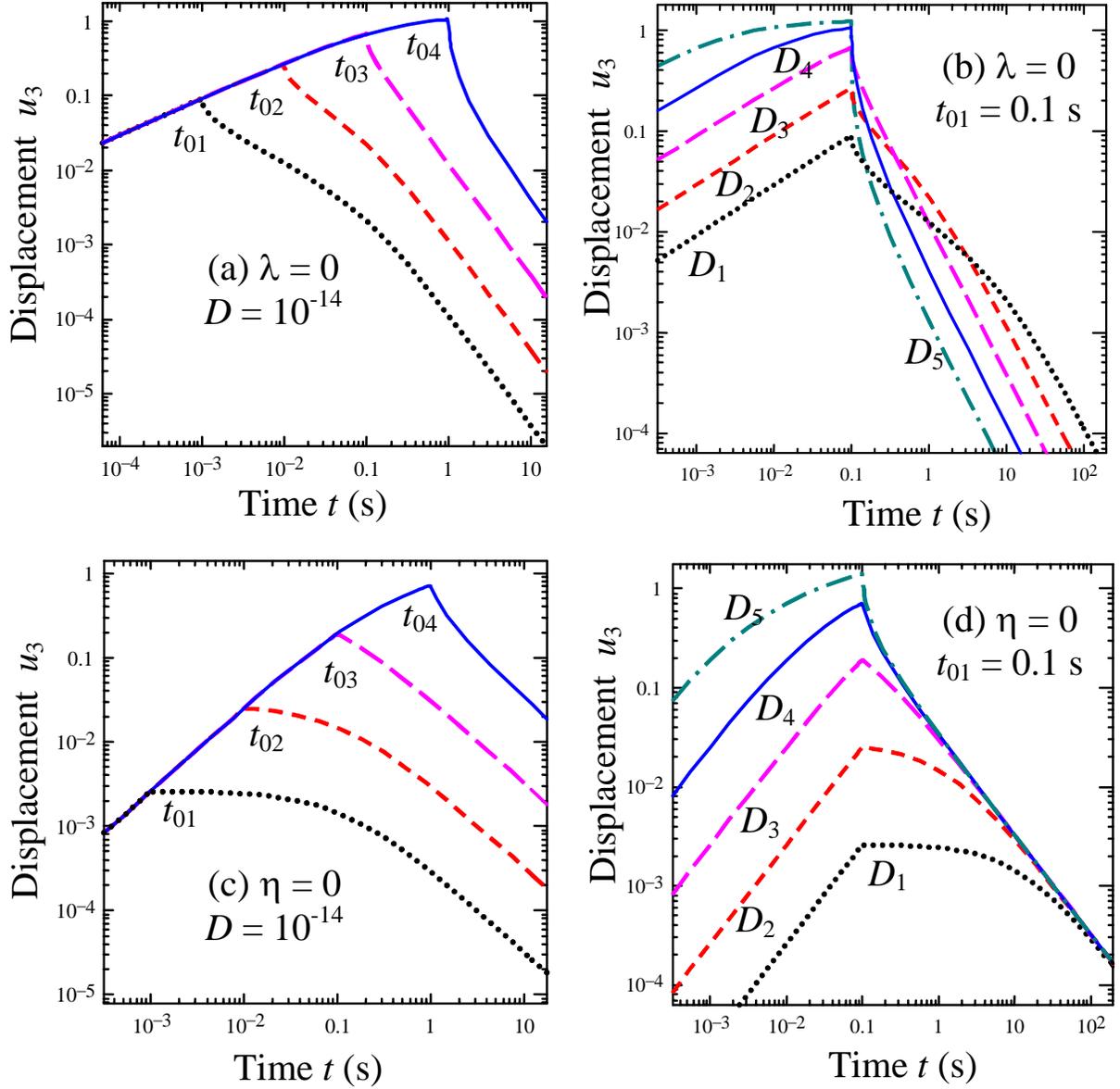

**Fig. 10.** Displacement $u_3(0,t)/\beta V_a$ relaxation in response to rectangular pulse for the case $\lambda=0$ (a,b) and $\eta=0$ (c,d), $R_0 = 100$ nm and (a,c) fixed diffusion coefficient $D = 10^{-14}$ m²/s and different pulse duration $t_0 = 10^{-3}, 10^{-2}, 10^{-1}, 1$ s; (b,d) fixed pulse duration $t_0 = 10^{-1}$ s and different diffusion coefficient $D = 10^{-16}, 10^{-15}, 10^{-14}, 10^{-13}, 10^{-12}$ m²/s.

Fig. 11 demonstrates the time dependence of the response $u_3(0,t)$ for different contact radius $R_0$. For the case of concentration-driven process ($\lambda=0$) the temporal relaxation at $t > t_0$ becomes logarithmic with $R_0$ value increase (compare plots a-c). For the case of flux-controlled process ($\eta=0$) the relaxation at $t > t_0$ demonstrates the broad plateau (which length increases with $R_0$ increase) and only then becomes logarithmic with time $t$ (compare plots d-f). Actually the case $R_0 > 10$ μm almost corresponds to the plain top electrode.

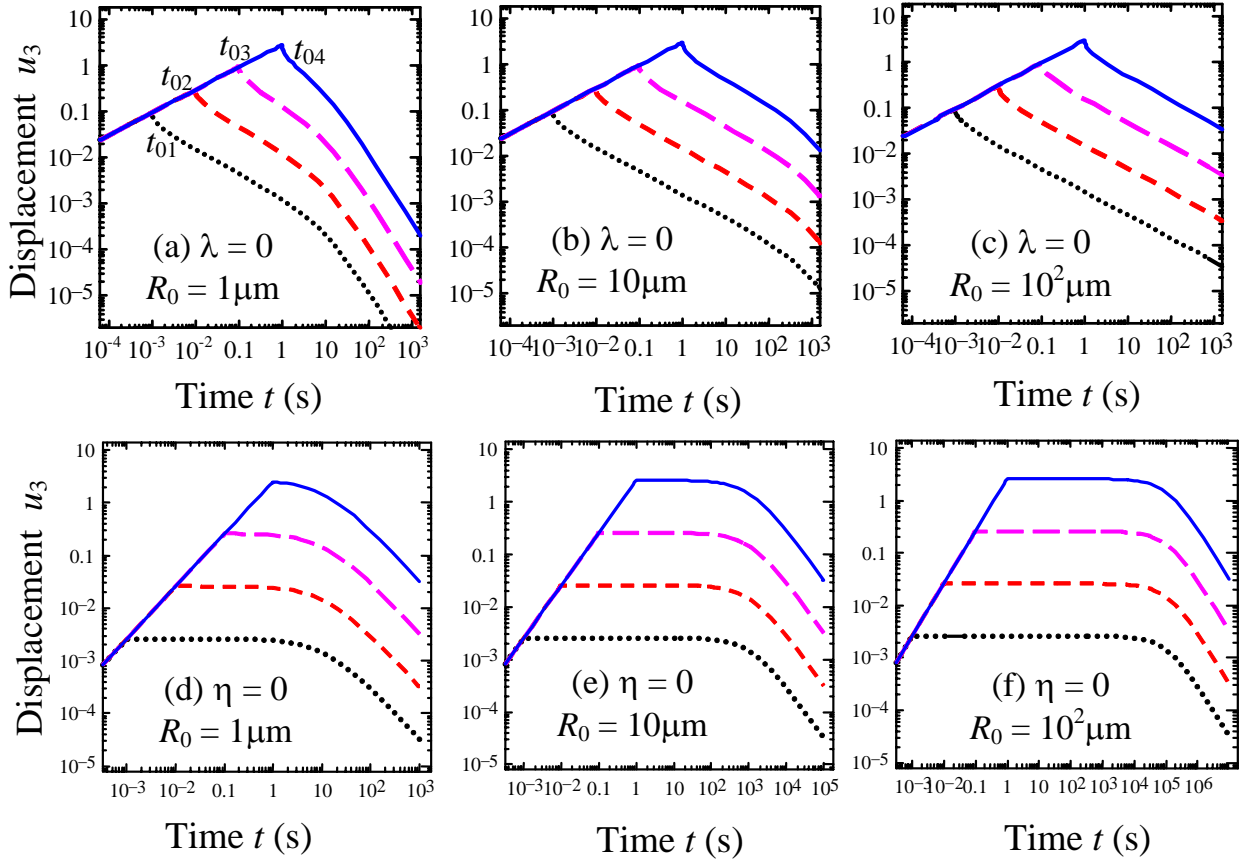

**Fig. 11.** Displacement $u_3(0,t)/\beta V_a$ relaxation in response to rectangular pulse for the case $\lambda=0$ (plots a-c) and $\eta=0$ (plots d-f), different values $R_0 = 1; 10; 100$ μm (see labels), fixed diffusion coefficient $D = 10^{-14}$ m$^2$/s and different pulse duration $t_0 = 10^{-3}$, $10^{-2}$, $10^{-1}$, 1 s (dotted, short-dashed, long-dashed and solid curves).

## V. Discussion
### V.1. Frequency regimes

In this section, we analyze the ESM signal for several realistic materials, and derive the sensitivity and imaging regimes that can be realized with modern scanning probe microscopy instrumentation. The characteristic length scale of Li-ion diffusion is given by the diffusion length, $L_D(\omega) = \sqrt{D_{Li}/\omega}$. The characteristic scale of the SPM detection is given by the tip-surface contact radius that ranges from 1-3 nm for ideal contact in the dry (or liquid) environment to 10-100 nm in ambient due to the tip wear and large capillary and electrocapillary tip-surface forces. Additionally, the size of the excitation volume can be larger then the area of mechanical contact due to the formation of water meniscus at the tip-surface junction. Finally, in liquids the characteristic size of excited volume is controlled by the frequency and conductive properties of medium, and only limited information for conventional un-insulated metal-coated probes is available.[29,30] For the insulated and shielded probes,[51,52,53,54] the length scale of electric field is given by the size of microelectrode [note that while insulated probes for STM type experiments can be readily prepared in-house, insulated AFM-type probes necessitate developed microfabrication process]. However, note that in spatially resolved imaging resolution, and hence the probe size, can often be determined self-consistently from the images, providing the self-consistent estimate on the characteristic tip size.

For frequency dependence of the diffusion length for typical diffusion coefficient of the order of $10^{-16} - 10^{-13}$ m$^2$s$^{-1}$ for Li$_x$CoO$_2$ ($x = 0.5 - 1$), LiMn$_2$O$_4$ and LiC$_6$ are shown in Fig. 12a. Note that the diffusion length reaches the length scale of 1 – 100 nm for times of the order of $10^{-2} - 10^2$ s. This time scales are well compatible with the SPM-based imaging (~10 ms/pixel) and spectroscopic imaging (100x100 pixel image in ~10 h, corresponding to ~3-4 s/pixel). At the same time, corresponding frequencies $10^2 - 10^{-2}$ Hz are well-below the characteristic resonant frequencies of the cantilever in the contact mode, ranging from ~10 kHz for soft cantilevers in liquids to ~0.2-0.7 MHz for stiff cantilevers used in contact mode imaging. Note that the imaging at cantilever resonances offers the advantage of mechanical amplification of the weak strain signal by the factor of 10-100 (i.e. *Q*-factor of the cantilever) and minimization of 1/*f* noise, and hence is of a special interest for SPM operation.

For comparison, Fig. 12 b shows the characteristic diffusion time $\tau_{Li}(R_0) = R_0^2/2D$ vs. the contact radius $R_0$ for different Li-containing materials Li$_x$CoO$_2$ ($x = 0.5 - 1$), LiMn$_2$O$_4$ and LiC$_6$. For radii from 1 to 100 nm the characteristic time is from $10^{-3} - 10^{-2}$ s to $1 - 10^2$ s depending on the material diffusion coefficient.

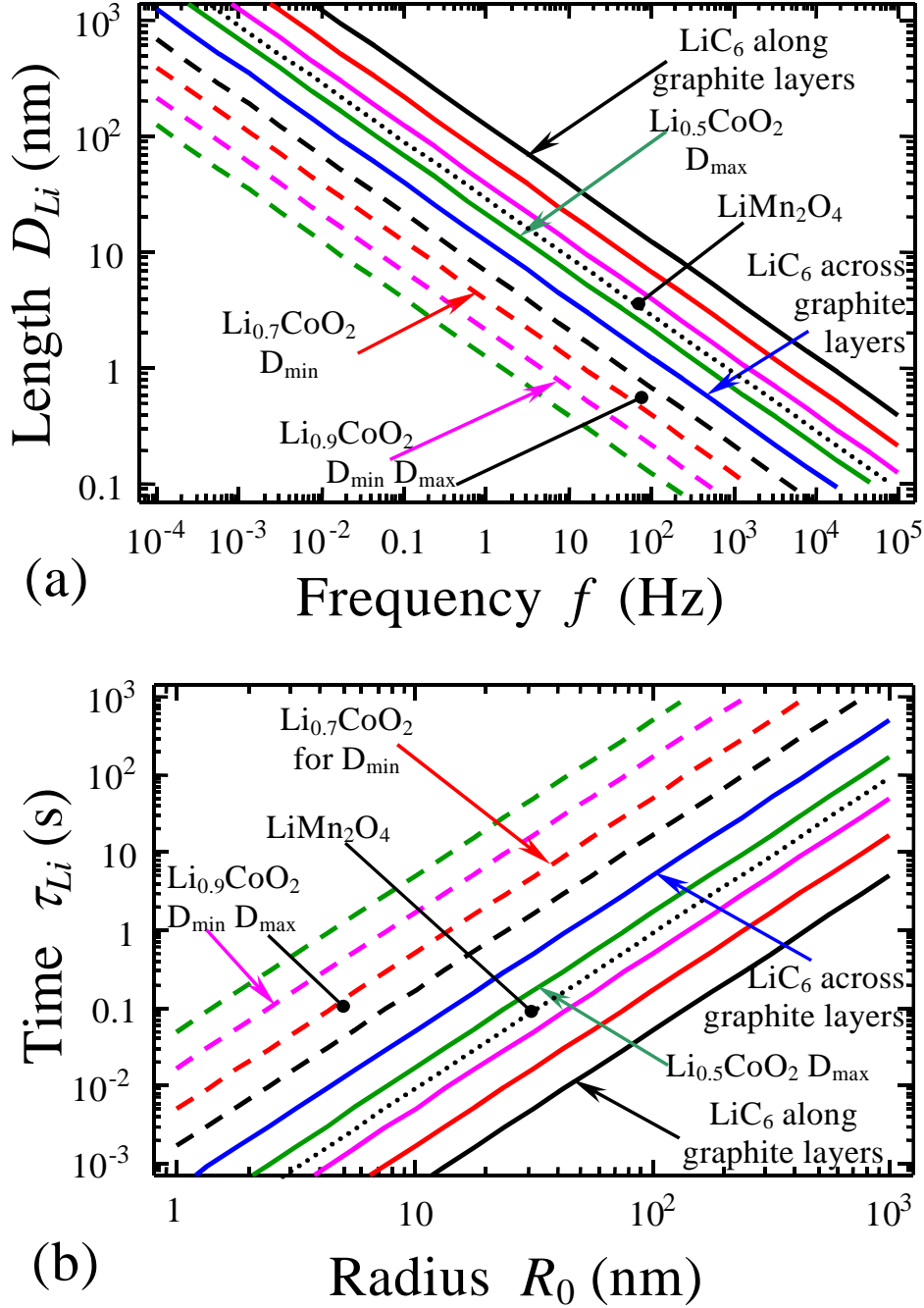

**Fig. 12.** (a) Diffusion length $L_D(\omega)$ vs. frequency $f = \omega/2\pi$ for different composition of $Li_xCoO_2$ ($x = 0.5 - 1$), $LiMn_2O_4$ and $LiC_6$ corresponding to diffusion coefficients $D_{Li}(x) = 7.1 \times 10^{-16} - 1 \times 10^{-13}$ m$^2$s$^{-1}$ (see curves from the top to bottom and Table 3). (b) Characteristic diffusion time $\tau_{Li}(R_0)$ vs. the contact radius $R_0$ for the same $D_{Li}$ as in plot (a).

**Table 3**. Diffusion coefficients, as a function of composition [55], specific molar expansion Vegard tensor $\beta_{ii}$. Lattice constants as a function of composition $x$ were taken from Refs.[56, 57], they correlate with results of Ref. [58].

| Composition | $D_{min}$ (m$^2$s$^{-1}$) | $D_{max}$ (m$^2$s$^{-1}$) | $\beta_{22} = \beta_{11} = \dfrac{1}{a}\dfrac{da}{dx}$ | $\beta_{33} = \dfrac{1}{c}\dfrac{dc}{dx}$ | $C_{max}$ (kmol/m$^3$) |
|---|---|---|---|---|---|
| Li$_{0.5}$CoO$_2$ | 7.1×10$^{-16}$ | 3.2×10$^{-15}$ | unknown | unknown | unknown |
| Li$_{0.6}$CoO$_2$ | 6.0×10$^{-16}$ | 2.8×10$^{-15}$ | ≈0 | 0.070 | unknown |
| Li$_{0.7}$CoO$_2$ | 1.0×10$^{-16}$ | 7.1×10$^{-16}$ | ≈0 | 0.049 | unknown |
| Li$_{0.8}$CoO$_2$ | 3.5×10$^{-17}$ | 2×10$^{-16}$ | ≈0 | 0.014 | unknown |
| Li$_{0.9}$CoO$_2$ | 3.0×10$^{-17}$ | 1.5×10$^{-16}$ | ≈0 | 0.035 | unknown |
| LiCoO$_2$ | 4.0×10$^{-17}$ | 2.5×10$^{-16}$ | ≈0 | 0.078 | 51.554 |
| LiMn$_2$O$_4$ (spinel) | unknown | 7.08×10$^{-15}$ | 0.027 | 0.027 | 22.900 |
| LiC$_6$ (Li-graphite) | 1.0×10$^{-15}$ across layers | 1.0×10$^{-13}$ along layers | 0.012 | 0.104 | 30.555 |

Based on these considerations, we distinguish three characteristic frequency regimes of the ESM imaging and spectroscopy:

**I. Static regime**, in which the driving frequency is below the inverse diffusion time and well below cantilever resonant frequency.

**II. Low frequency regime**, in which the driving frequency is larger then the inverse diffusion time, but is well-below the cantilever resonance frequency.

**III. Dynamic (or high-frequency) regime**, in which the driving frequency is well above the inverse diffusion time and is comparable to the resonant frequency of the cantilever.

Below, we briefly discuss the expected aspects of response in these limits.

### V.1.1. Response in static and low-frequency regimes

The relationship between the amplitude of surface vibrations and measured cantilever deflection is given by the transfer function. The latter is dominated by the cantilever dynamics in the Hz-Mhz ranges, while at higher frequencies the photodetector and electronic behavior becomes significant. Here, we discuss cantilever dynamics effects only, since the other effects are a part of the measurement system and can be established and calibrated independently.

In the static and low-frequency regimes, the transfer function of the SPM cantilever is (ideally) linear. Practically, the response function exhibits significant dispersion below 10 kHz due to multiple (typically narrow) tip-holder resonances. However, these are generally unrelated to the variation of mechanical properties of the tip-surface junction, and hence do not result in topographic cross-talk (since transfer function at each frequency can be calibrated). Thus, the frequency dependence of the ESM response is effectively controlled by the Li-ion dynamics.

In the static regime, the Li-diffusion length significantly exceeds the signal detection volume of the SPM probe, i.e. the bias-induced changes occur on the length scale larger then SPM probe size. In the limit of dc bias ramp, electrochemical transformation can occur in the full volume of the material, with the possible onset of irreversible dynamics and limits on the number of charge-discharge cycles. The advantage of this regime is that the surface displacement can be measured quantitatively (albeit note that due to uncertainty in the tip size and relationship between probe potential and Li-concentration change, this information may be insufficient for characterization of electrochemical functionality).

In the low frequency regime, the Li-diffusion length is comparable to the probe size, allowing for the optimal compromise between the potential to study process in detail (i.e. control the degree of lithiation by changing the voltage range) and bandwidth of measurements (i.e. the spatially resolved measurements are possible). Furthermore, the small area affected by electrochemical process increases the degree of reversibility of the process. The transfer function of the SPM in this frequency interval is ideally constant, i.e. measured response will be defined only by Li-ion dynamics. Practically, the transfer function has significant dispersion (esp. below 10 kHz) due to the tip-holder resonances. However, these are position independent, and result only in the change of the absolute value of response signal and can be calibrated. Fig. 13 shows the frequency dependence of the electromechanical response real, imaginary part, absolute value and phase for several Li-containing materials $Li_xCoO_2$, $LiMn_2O_4$ and $LiC_6$. In all cases, we assume that that driving voltage is small, so that the change in Li concentration is small, the diffusion coefficients are constant and linear theory is applicable.

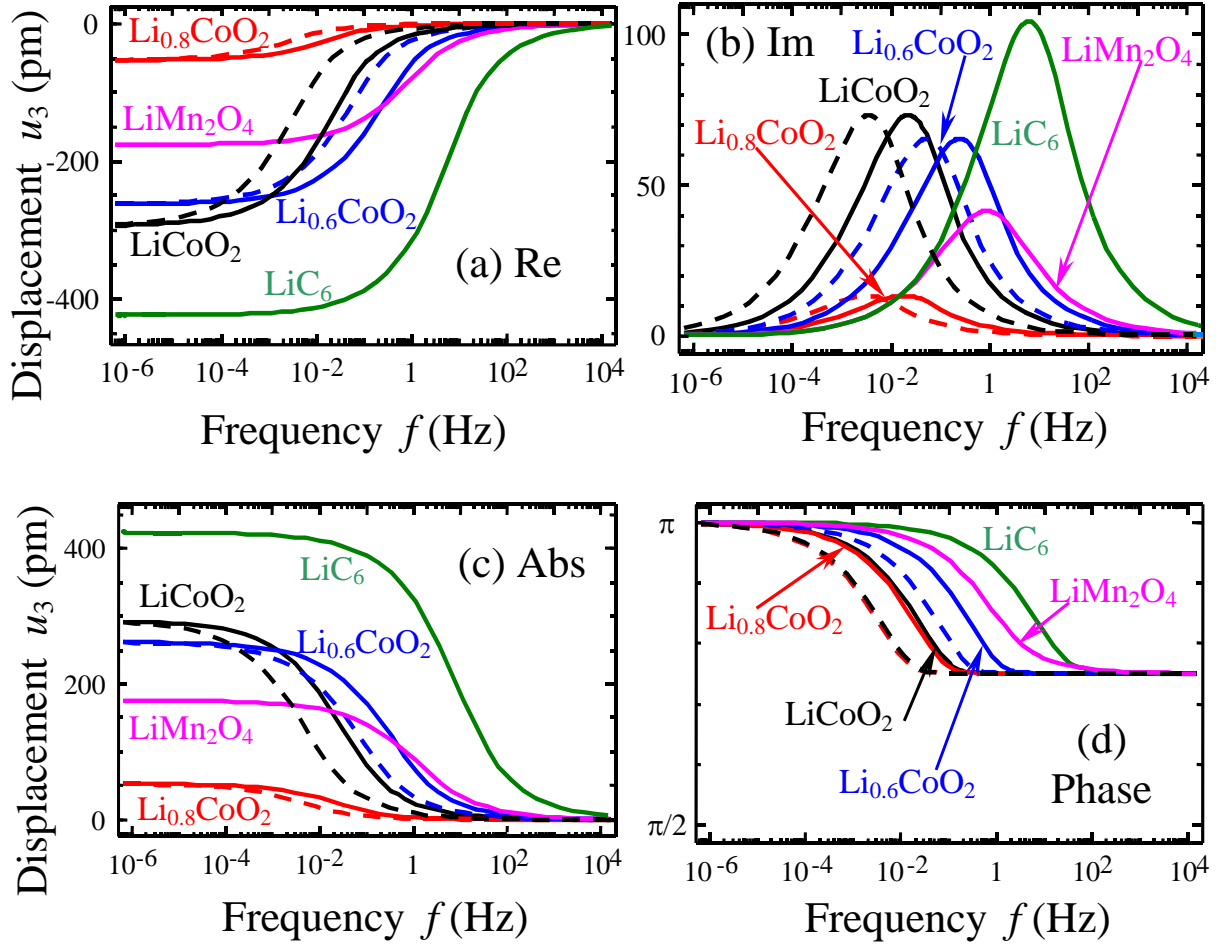

**Fig. 13.** Displacement $u_3(0,w)$ real (a) and imaginary (b) parts; absolute value (c) and phase (d) vs. driving frequency $f$ for determined concentration $\delta C/C_{max} = 0.05$ (i.e. $\lambda=0$) for different Li-containing materials $Li_xCoO_2$ with $D_{min}$ (dashed curves) and $D_{max}$ (solid curves), $LiMn_2O_4$ and $LiC_6$ (labels near the curves). Diffusion coefficient range $D_{Li}(x)$ Vegard tensor $\beta_{ii}$ and maximal lithium concentration $C_{max}$ are listed in the Table 3. Note that $LiMn_2O_4$ is isotropic material with respect to Vegard tensors, other ones are anisotropic.

Despite the fairly character of approximations in this section, the estimated response amplitudes are of the order of ~0.05 – 0.2 nm even for small ($\delta C/C_{Li} = 0.05$) variations of Li-concentration. These values are well within theoretical limits of SPM detection (in the absence of 1/f noise). Furthermore, it can be expected that application of sufficiently high driving biases (corresponding to $\delta C/C_{Li} \sim 1$) can yield surface deformations in the 1 – 4 nm range, well-detectable as an SPM strain loop in the ~1-10 s acquisition times.[59]

Fig. 14 shows the frequency dependence of the concentration flux real and imaginary parts for different Li-containing materials for determined concentration $\delta C/C_{Li} = 0.05$.

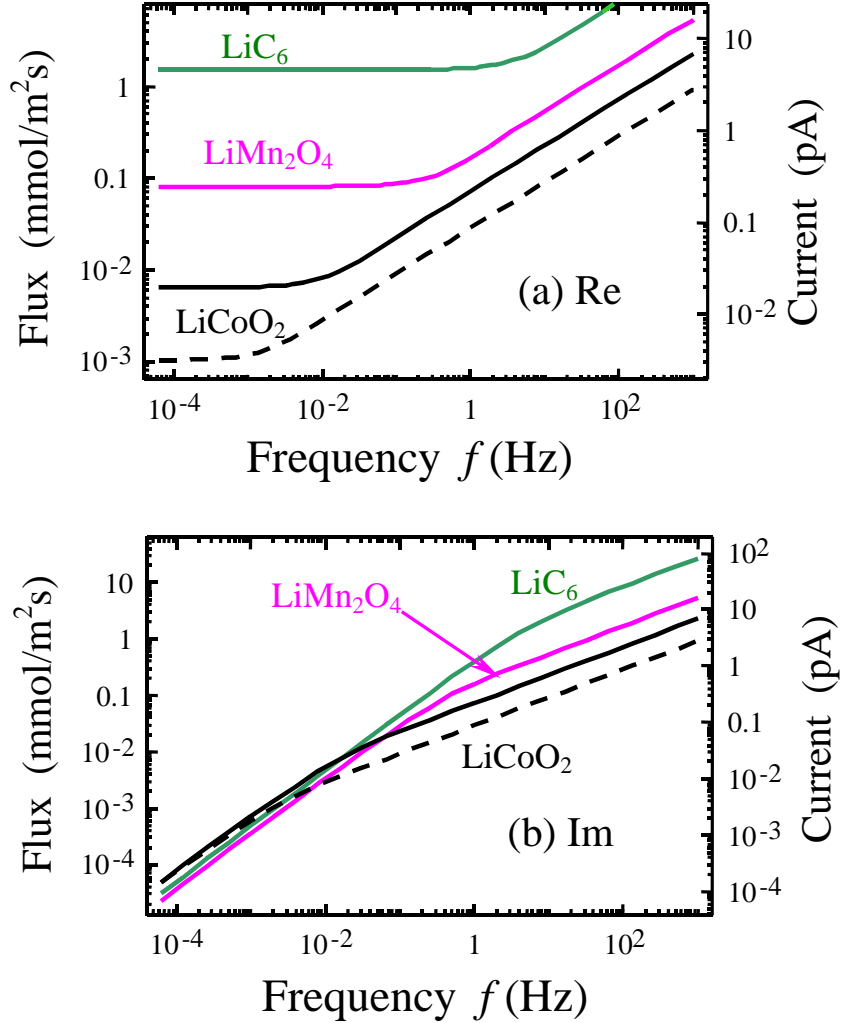

**Fig. 14.** Concentration flux real (a) and imaginary (b) parts vs. driving frequency $f$ for determined concentration $\delta C/C_{max} = 0.05$ (i.e. $\lambda=0$), $R_0 = 100$ nm and different Li-containing materials $LiCoO_2$, $LiMn_2O_4$ and $LiC_6$ (labels near the curves). Diffusion coefficients $D_{Li}(x)$, tensor $\beta_{ii}$ and $C_{max}$ are listed in the Table 3. Total electric current was estimated as $I_c \approx -\pi R_0^2 e Z_c D(\partial \delta C/\partial z)$ (see right scale).

### V.1.2 Imaging in the high frequency regime.

In the high frequency regime, the excitation frequency is comparable to the resonance frequency of the cantilever (0.03 – 1 MHz) and well above the inverse diffusion time of lithium even on the length-scales of 1 -30 nm (i.e. characteristic probe size of SPM tip). Correspondingly, even high driving forces will result only in minute changes in Li concentration, as analyzed in Section III.2. Note that the applicability of continual models on the length scales below Li diffusion length presents an interesting problem not addressed in this manuscript.

The measured SPM signal in the absence of electrostatic tip surface forces, etc. is given by the product of the surface displacement Eq. (4) and tip-surface transfer function. For the electromechanically driven SPM, the tip-surface contrast transfer[60,61,62] and contribution of electrostatic forces to signal[63,64] was extensively studied in the context of Piezoresponse Force Microscopy of piezoelectric and ferroelectric materials. Similarly, cantilever dynamics including the dissipative and damping effects has been explored in the context of Atomic Force Acoustic Microscopy.[65,66] Here, we illustrate the ESM response in the high-frequency regime using experimental transfer function for the stiff (5 N/m) cantilever measured with acoustic excitation. For simplicity, we estimate the response as the product of the respective amplitudes and ignore the phase content of the signal. This assumption is well justified, since the regions with large frequency dispersion for the materials and cantilever response are well separated in frequency.

Shown in Figs. 15 a,b are the materials response and ESM signal in the 10 mHz – 10 MHz range. Note that while the intrinsic materials response rapidly decays above ~ 1 Hz, this decay is relative slow power of frequency (~$1/f$). Correspondingly, the presence of sharp (Q ~100) cantilever resonances results in the strong amplification of the ESM signal. Directly at the resonance, the high-frequency ESM signal is comparable to that in the static regime.

The advantageous features of high-frequency ESM can be illustrated once the sources of noise in the SPM detection system are taken into account.[67] For sufficiently high frequencies, the ultimate noise level of SPM is given by the thermomechanical white noise (practically, most commercial AFMs with appropriate vibrational isolation operate within 1-1.5 order of magnitude from this limit). However, for low frequencies the noise is $1/f$ character, in which case averaging the signal over large times does not allow more precise measurements. The $1/f$ noise corner is typically in the 1-10 kHz range. To estimate the noise effect on ESM signal, plotted in Fig. 16d is the signal/noise ratio, with the noise being approximated as $N = (f_0/f+1)$. Notice that while $1/f$ noise strongly affects the static and low-frequency ESM measurements, the high-frequency ESM allows imaging with high signal-noise ratios despite that only small changes of Li concentration can be induced.

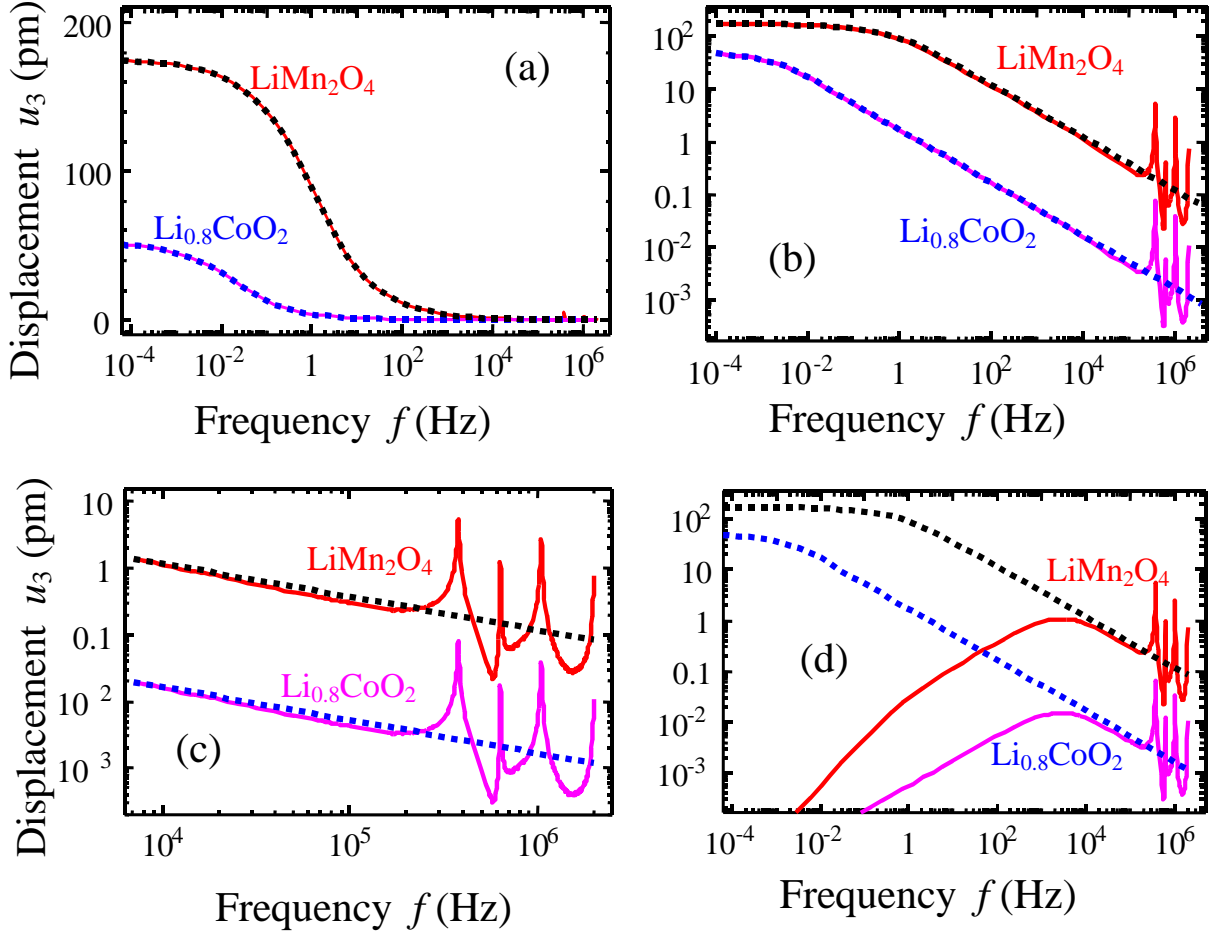

**Fig. 15.** (a,b) Absolute value of the material response (displacement $u_3(0,w)$, dotted curves); and the ESM response ($u_3(0,w)$ multiplied by experimental transfer function, solid curves) vs. driving frequency $f$ for determined concentration $\delta C/C_{max} = 0.05$ (i.e. $\lambda=0$) for LiMn$_2$O$_4$ and Li$_{0.8}$CoO$_2$ (labels near the curves). (c) Zoom in the high-frequency regime. (d) Material response (dotted curves) and signal to noise ratio for ESM response divided by noise function $\sim (f_0/f+1)$ with $f_0$=3 kHz (solid curves). Diffusion coefficients $D_{Li}(x)$, tensor $\beta_{ii}$ and $C_{max}$ are listed in the Table 3.

Note that the use of high-frequency regimes necessitates the use of the resonant-frequency tracking methods, since cantilever resonance frequency is strongly dependent on surface topography through the tip-surface spring constant.[68,69] Similarly to PFM, the use of standard phase-locked loop circuitry for ESM is unlikely to succeed, since the phase of the local electromechanically response is dependent on materials functionality, and is hence not constant (unlike SPMs with acoustic excitation, for which this condition is semi-quantitatively satisfied). Correspondingly, the use of amplitude-based resonance tracking[70] or band excitation method is required.[71]

### V.2. The role of electrochemical reactions and limitations

The linear theory developed in Section II-IV describes the signal formation mechanism in ESM in the linear limit, in which the change in Li concentration is relatively small. Here, we qualitatively consider the effects of finite Li concentration and electrochemical reaction on responses.

The saturation effect is related to the limit of Li concentration in real materials. In this case, the concentration cannot exceed the intercalation limit, and change in Li concentration induced by progressively high bias amplitude will saturate, as shown in Fig. 16 a. The expected shape of the hysteresis loop in this case is shown in Fig. 16 b.

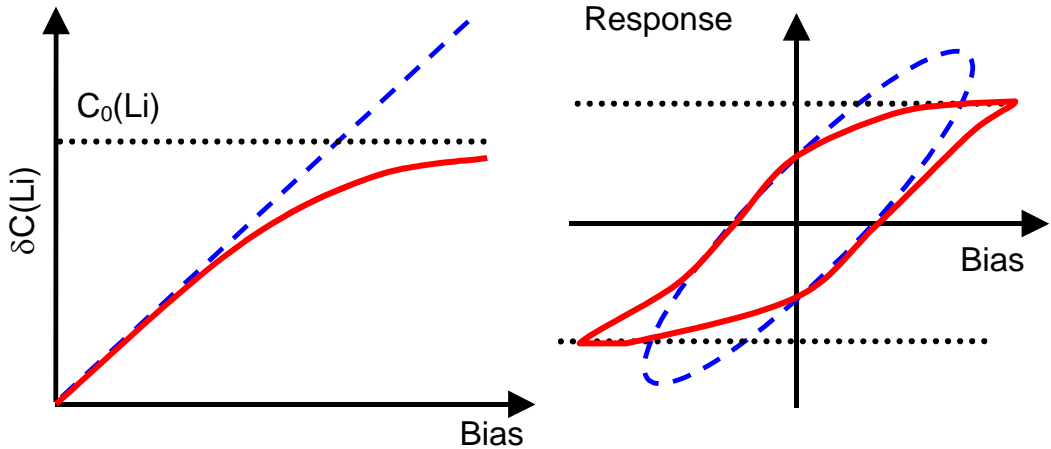

**Fig. 16**. (a) Bias dependence of the lithium concentration change for ideal linear case (blue dashed line) and realistic case (red curve). (b) Expected evolution of the hysteresis loops. Ideal

case (blue dashed loop) and realistic case (red solid loop). (*Sorry for conversion, but I cannot edit Canvas and insert it into the text as editors demand, so let all figs be inserted metafiles*)

The second factor potentially affecting the ESM signal formation mechanism is the electrochemical reaction in the tip-surface junction. In this case, the Li insertion does not start until the critical overpotential is achieved. We note that for most processes the process is reaction limited for small overpotentials, and diffusion limited by large overpotentials due to exponential Tafel-like dependence of reaction rate on the overpotential. The width of transition is typically of the order of ~0.1-0.3 V.[72] Correspondingly, the response as a function of bias and hysteresis loops can be expected to behave as shown in Fig. 17. Note that the fine shape of hysteresis loop at sufficient resolution may provide the information on the phase transition sequence in the material, providing electromechanical analog of cyclic voltammetry.

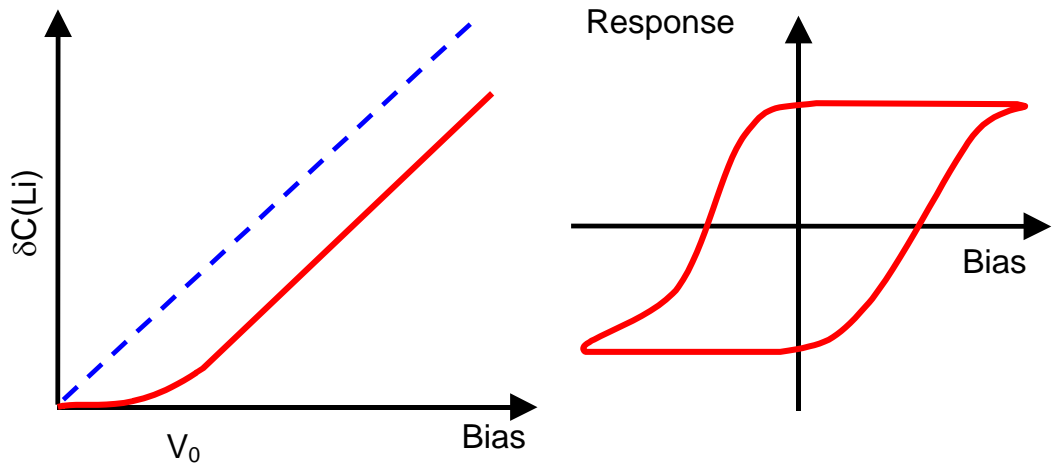

**Fig. 17**. (a) Bias dependence of the lithium concentration change for ideal linear case (dashed blue line) and in the presence of limiting electrochemical reaction stage (solid red curve). (b) Expected shape of the ESM hysteresis loops.

### V.3. Sensitivity limits for ESM

One of the interesting issues in ESM is the ultimate resolution limit of the ESM. The simple estimate illustrates that vertical displacement of the material surface due to intercalation can be detected down to the nanometer scales, as a consequence of extremely high vertical sensitivity of AFM (note that even in the static regime the work by Shao-Horn group has demonstrated direct measurements of step height in $LiCoO_2$ as a function of intercalation degree). Here, we estimate the resolution limit for the ESM, following the detailed analysis for PFM reported in Ref. [73].

Briefly, in the absence of electrostatic forces the detection limit in PFM is set by the signal transfer from material to the tip, determined by the cantilever spring constant, $k$, and tip-surface spring constant, $k_1$. The contact stiffness $k_1 = (\partial h/\partial P)^{-1}$, and in the simple Hertzian model can be found as $k_1 = 2aY^*$, or

$$k_1 = 2Y^*\sqrt{hR_0} = \left(6PY^{*2}R_0\right)^{\frac{1}{3}}, \tag{20}$$

where $h$ is the indentation depth, $R_0$ is the tip radius of curvature, and $P$ is load.[42] The $Y^*$ is the effective Young's modulus of the tip-surface system. The contact radius, $a$, is related to the indentation depth as $a = \sqrt{hR_0}$. Notably, for typical cantilever spring constant $k$ = 1 – 40 N/m, the condition $k_1 > k$ is satisfied for $a$ > 0.01 – 0.4 nm, i.e. practically for all feasible contact areas. Correspondingly, the limiting factor in ESM resolution is not the contact transfer, but rather the minimal tip-surface forces as limited by adhesion or capillary forces (in ambient). The latter are typically of the order of 100 nN in ambient environment, corresponding to the spatial resolution of order of 3 – 10 nm.

The second limiting factor in resolution is the contribution of the electrostatic forces to the signal. While (theoretically) non-hysteretic, large additional force contribution results in increase of the force noise and hysteretic phenomena due to electrocapillary[74] interactions and instrumental transients. The condition for the dominance of the electromechanical interactions was derived as $a > a^* = C'_{sphere}(V_{dc} - V_s)/2u_3Y^*$, where $a^*$ is the critical contact radius corresponding to equality of the electrostatic and electromechanical contributions to the signal, $V_{dc}$ is applied $dc$ bias, $V_s$ is the static surface potential bias. For prototypical cathode material (100 GPa, 20 pm/V), this conditions becomes $a > a^* = 12.5(V_{dc} - V_s)$ A/V. From this simple estimation, the resolution of low-frequency ESM on hard materials can potentially achieve sub-10 nanometer scale, provided that the electrostatic contribution to the signal is minimized.

Note that the key aspect of the electromechanical detection principle in ESM, as opposed to the current detection in conventional electrochemical characterization techniques is that the electronic transport does not directly contribute to strain. Hence, electrochemically induced strains provide the information on ion dynamics even in the presence of large electronic currents, and hence are provide relevant information even at high frequencies and small contact areas.

## VI. Summary

The image formation mechanism in the electrochemical strain microscopy is analyzed. The ESM utilizes the strong coupling between ionic concentration and strains in Li-ion conductors, to deduce the information on ionic flow from oscillatory mechanical surface responses. This approach allows effectively separating ionic currents and electronic currents, and hence allows high-veracity measurements of the former. The extremely high sensitivity of modern scanning probe microscopies achieving 3-10 pm in the 1 kHz bandwidth allows measurements of strain-coupled electrochemical processes on the sub-10 nm levels. The response signals in high-frequency EPFM are analyzed, as are time dynamics, providing the local analogs of the conventional current-based electrochemical methods. We believe that future development of the method will allow mapping of kinetics of thermodynamics of electrochemical reactions in solids on the nanometer level of individual grains and ultimately structural defects, providing much-needed knowledge on nanoscale mechanisms underpinning battery functionality.

This material is based upon work supported as part of the Fluid Interface Reactions, Structures and Transport (FIRST) Center, an Energy Frontier Research Center funded by the U.S. Department of Energy, Office of Science, Office of Basic Energy Sciences under Award Number ERKCC61 (N.B., S.V.K.). Research at the ORNL's Center for Nanophase Materials Sciences in the project CNMS2010-098 and CNMS2010-099 was sponsored by the Scientific User Facilities Division, Office of Basic Energy Sciences, U. S. Department of Energy (N.B.). N.B. acknowledges the Alexander von Humboldt foundation for financial support. ANM and EEA gratefully acknowledge financial support from National Academy of Science of Ukraine, Ministry of Science and Education of Ukraine (UU30/004) and National Science Foundation (DMR-0908718). SVK and NB gratefully acknowledge S.J. Pennycook and R.E. Garcia for valuable discussions.

## Appendix A. Elastic problem solution

For the quasi-static case, the equations of state $\delta_{ij}\beta\delta C + s_{ijkl}\sigma_{kl} = u_{ij}$ for isotropic media can be rewritten as:[43]

$$u_{11} = s_{11}\sigma_{11} + s_{12}(\sigma_{22} + \sigma_{33}) + \beta_{11}\delta C, \qquad u_{22} = s_{11}\sigma_{22} + s_{12}(\sigma_{11} + \sigma_{33}) + \beta_{22}\delta C \qquad (A.1)$$

$$u_{33} + s_{11}\sigma_{33} + s_{12}(\sigma_{22} + \sigma_{11}) + \beta_{33}\delta C, \qquad (A.2)$$

$$u_{12} = (s_{11} - s_{12})\sigma_{12}, \quad u_{13} = (s_{11} - s_{12})\sigma_{13}, \quad u_{23} = (s_{11} - s_{12})\sigma_{23}. \qquad (A.3)$$

Here δC is the concentration redistribution.

Since we are interested in solution in terms of displacement, let us write the conditions, which strain and stress distribution should satisfy, namely compatibility condition

$$\text{inc}(i,j,\hat{u}) = e_{ikl}e_{jmn}u_{ln,km} = 0 \qquad (A.4)$$

and equilibrium conditions, both in the bulk and on the free surface

$$\sigma_{ij,j} = 0, \qquad \sigma_{ij}n_j\big|_S = 0 \qquad (A.5)$$

Here comma separated subscript means the derivative on corresponding coordinate, e.g. $\partial\sigma_{ij}/\partial x_k \equiv \sigma_{ij,k}$.

Using the definition of strain components $u_{ij} = (u_{i,j} + u_{j,i})/2$ and the equation of state in the form $\sigma_{kl} = c_{klij}u_{ij} - c_{klii}\beta\delta C$, it is easy to get the following equation determining the distribution of displacement vector

$$\sigma_{ij,j} = 0 \quad \Rightarrow \quad c_{ijkl}u_{k,lj} - c_{ijkk}\beta\delta C_{,j} = 0 \qquad (A.6)$$

and corresponding boundary conditions

$$\sigma_{ij}n_j\big|_S = 0 \quad \Rightarrow \quad (c_{ijkl}u_{k,l} - c_{ijkk}\beta\delta C)n_j\big|_S = 0 \qquad (A.7)$$

Using the general equation of equilibrium, written in terms of displacement vector **u** is

$$\frac{Y}{2(1+\nu)}\left(\Delta_x \mathbf{u} + \frac{1}{1-2\nu}\text{grad}_x(\text{div}_x \mathbf{u})\right) + \mathbf{F}\cdot\delta(\mathbf{x}-\boldsymbol{\xi}) = 0 \qquad (A.8)$$

where vector **x** denotes the point where we look for the solution, $\boldsymbol{\xi}$ is the point were point force **F** is applied. ν is Poisson coefficient, Y is Young modulus (so that $c_{ijkl} = \frac{Y}{2(1+\nu)}\left(\frac{2\nu}{1-2\nu}\delta_{ij}\delta_{kl} + \delta_{ik}\delta_{jl} + \delta_{il}\delta_{jk}\right)$), one can see that the considered case of equations (A.6) and Eq.(A.7) can be considered as the usual problem of the elasticity theory with bulk force density $-c_{ijkk}\beta_{kk}\delta C_{,j}$ and surface force density, acting in the elastic media (or/and on its surface) $c_{ijkk}\beta_{kk}\delta C\, n_j$.

In order to present the solution of Eq.(A.8) for the elastic semi-space with a free surface, it is convenient to use 2D-Fourier transformation

$$\tilde{u}_i(k_1,k_2,x_3) = \frac{1}{2\pi}\int_{-\infty}^{\infty}dx_1\int_{-\infty}^{\infty}dx_2\,\exp(ik_1x_1+ik_2x_2)\cdot u_i(\mathbf{x}), \qquad (A.9a)$$

$$u_i(\mathbf{x}) = \frac{1}{2\pi}\int_{-\infty}^{\infty}dk_1\int_{-\infty}^{\infty}dk_2\,\exp(-ik_1x_1-ik_2x_2)\,\tilde{u}_i(k_1,k_2,x_3), \qquad (A.9b)$$

where $k_1$ and $k_2$ are the components of 2D wave vector $\mathbf{k}$ with the module $k=\sqrt{k_1^2+k_2^2}$.

The Fourier image of the solution of Eq.(A.8) can be represented as

$$\tilde{u}_i^s(k_1,k_2,x_3) = \tilde{G}_{ij}^s(k_1,k_2,x_3,\xi_3)F_j\exp(ik_1\xi_1+ik_2\xi_2). \qquad (A.10a)$$

and distribution in r-space

$$u_i^s(\mathbf{x}) = \frac{1}{2\pi}\int_{-\infty}^{\infty}dk_1\int_{-\infty}^{\infty}dk_2\,\exp(-ik_1(x_1-\xi_1)-ik_2(x_2-\xi_2))\,\tilde{G}_{ij}^s(k_1,k_2,x_3,\xi_3)F_j \equiv$$
$$\equiv G_{ij}^s(x_1-\xi_1,x_2-\xi_2,x_3,\xi_3)F_j \qquad (A.10b)$$

Here $G_{ij}^s(x_1-\xi_1,x_2-\xi_2,x_3,\xi_3)$ is the Green's function of the corresponding problem for point force.

Using the Green's function $u_i(\mathbf{x})=\iiint\limits_{\xi_3>0}G_{ij}^s(x_1-\xi_1,x_2-\xi_2,x_3,\xi_3)f_j(\xi_1,\xi_2,\xi_3)d\xi_1 d\xi_2 d\xi_3$, here $f_j(x_1,x_2,x_3)$ is the spatially distributed force. For the case of the effective forces with bulk density $f_j(x_1,x_2,x_3)=-c_{ijkk}\beta\,\delta C_{,j}$ and surface density $f_{jS}(x_1,x_2,x_3)=c_{ijkk}\beta\,\delta C\, n_j$ we obtained

$$u_i(\mathbf{x}) = -\iiint\limits_{\xi_3>0}G_{ij}^s(x_1-\xi_1,x_2-\xi_2,x_3,\xi_3)c_{jmkk}\,\beta\,\delta C(\xi_1,\xi_2,\xi_3)_{,m}d\xi_1 d\xi_2 d\xi_3 +$$
$$+ \iint\limits_{\xi_3=0}G_{ij}^s(x_1-\xi_1,x_2-\xi_2,x_3,0)c_{jmkk}\,\beta\,\delta C(\xi_1,\xi_2,0)n_m d\xi_1 d\xi_2 \qquad (A.11)$$

Elementary transformations and Gauss-Ostrogradsky lead to

$$u_i(\mathbf{x}) = \iiint\limits_{\xi_3>0}\frac{\partial G_{ij}^s(x_1-\xi_1,x_2-\xi_2,x_3,\xi_3)}{\partial \xi_m}c_{jmkk}\,\beta\,\delta C(\xi_1,\xi_2,\xi_3)d\xi_1 d\xi_2 d\xi_3 \qquad (A.12)$$

The elastically isotropic semi-space Green's function at $x_3 = 0$ is

$$G_{ij}^S(x_1-\xi_1, x_2-\xi_2, \xi_3) = \begin{cases} \frac{1+\nu}{2\pi Y}\left[\frac{\delta_{ij}}{R} + \frac{(x_i-\xi_i)(x_j-\xi_j)}{R^3} + \frac{1-2\nu}{R+\xi_3}\left(\delta_{ij} - \frac{(x_i-\xi_i)(x_j-\xi_j)}{R(R+\xi_3)}\right)\right] & i,j \neq 3 \\ \frac{(1+\nu)(x_i-\xi_i)}{2\pi Y}\left(\frac{-\xi_3}{R^3} - \frac{(1-2\nu)}{R(R+\xi_3)}\right) & i=1,2;\ j=3 \\ \frac{(1+\nu)(x_j-\xi_j)}{2\pi Y}\left(\frac{-\xi_3}{R^3} + \frac{(1-2\nu)}{R(R+\xi_3)}\right) & j=1,2;\ i=3 \\ \frac{1+\nu}{2\pi Y}\left(\frac{2(1-\nu)}{R} + \frac{\xi_3^2}{R^3}\right) & i=j=3 \end{cases}$$

(A.13)

Here $R = \sqrt{(x_1-\xi_1)^2 + (x_2-\xi_2)^2 + \xi_3^2}$, $\nu$ is the Poisson coefficient, $Y$ is the Young modulus (so that $c_{ijkl} = \frac{Y}{2(1+\nu)}\left(\frac{2\nu}{1-2\nu}\delta_{ij}\delta_{kl} + \delta_{ik}\delta_{jl} + \delta_{il}\delta_{jk}\right)$).

Mechanical displacement (7) is:

$$u_i(r,\omega) = -2(1+\nu)\beta\int_0^\infty d\xi_3 \int_0^\infty dk\, \exp\left(-\left(k + \sqrt{k^2 + i\omega/D}\right)\xi_3\right) \frac{R_0 V(\omega) J_1(kR_0)}{\lambda\sqrt{k^2+i\omega/D} - \eta} J_0(kr). \quad (A.14)$$

Integration on $\xi_3$ leads to

$$u_i(r,\omega) = \int_0^\infty dk\, \frac{-2(1+\nu)\beta V(\omega)\cdot R_0 J_1(kR_0) J_0(kr)}{\left(k + \sqrt{k^2 + \frac{i\omega}{D}}\right)\left(\lambda\sqrt{k^2 + \frac{i\omega}{D}} + \eta\right)} = \int_0^\infty dq\, \frac{-2(1+\nu)\beta V(\omega)\cdot R_0^2 J_1(q) J_0(qr/R_0)}{\left(q + \sqrt{q^2 + \frac{R_0^2}{D}i\omega}\right)\left(\lambda R_0\sqrt{q^2 + \frac{R_0^2}{D}i\omega} + \eta\right)}$$

(A.15)

## Appendix B

Electrostatic potential $V_e(\mathbf{x},t)$ distribution is found self-consistently from electrostatic equations.

$$\begin{cases} \left(\frac{\partial^2}{\partial x_1^2} + \frac{\partial^2}{\partial x_2^2} + \frac{\partial^2}{\partial x_3^2}\right) V_e(\mathbf{x},t) - \frac{V_e(\mathbf{x},t)}{R_d^2} = -\frac{\rho(\mathbf{x},t)}{\varepsilon_0 \varepsilon}, \\ \frac{\partial}{\partial t}\rho(\mathbf{x},t) = -\mathrm{div}(J_c(\mathbf{x},t) + J_a(\mathbf{x},t)) = 0. \end{cases}$$

(B.1)

Where the boundary conditions are $V_e(x_1, x_2, x_3 = 0, t) = V_0(x_1, x_2, t)$ at the tip electrode $x_3 = 0$ and $V_e(x_1, x_2, x_3 = h) = 0$ at remote bottom electrode $x_3 = h$.

Here, $R_d$ is the Debye screening radius, $\varepsilon_0$ is the dielectric constant, $\varepsilon$ is relative dielectric permittivity, $e$ is the absolute value of the electron charge, $h$ is the cell thickness. Electric charge density $\rho(\mathbf{x}, t) = eZ_c \delta C_c(\mathbf{x}, t) - eZ_a \delta C_a(\mathbf{x}, t)$ and electric current densities of the anions and cations are

$$J_c = -eZ_c \left( D_c \text{grad}\delta C_c + \mu_c \delta C_c \text{grad} V_e - \frac{C_c}{RT} \text{grad}(\beta_{ij}\sigma_{ij}) \right), \quad (B.2a)$$

$$J_a = -eZ_a \left( D_a \text{grad}\delta C_a - \mu_a \delta C_a \text{grad} V_e - \frac{C_a}{RT} \text{grad}(\beta_{ij}\sigma_{ij}) \right) \quad (B.2b)$$

$Z_{a,c}$ are relative charges (in the units of the electron charge) of the anions and cations correspondingly. $\delta C_{a,c}$ are anion and cation concentration excess, $D_{a,c}$ are anion and cation diffusion coefficients, $\mu_{a,c}$ are their mobilities.

When the current density at the tip electrode $x_3 = 0$ appeared only due to the cations ($J(\mathbf{x}, t) = J_c$, $J_a = 0$), a local deviation from equilibrium of the surface electrostatic potential, i.e. the overpotential $U_e(x_1, x_2, t)$, constitutes the driving force for the reaction to take place. For lithium such reaction is given by equation $Li^+ + e^- \leftrightarrow Li$. The rate of the ions transfer from the electrolyte to the active material phase is controlled by the Butler-Volmer relation [45]. Thus the total charge flux density [46]:

$$J_c(x_1, x_2, x_3 = 0, t) = i_0 \left( \exp\left(-\alpha_a \frac{FU_e}{RT}\right) - \exp\left(\alpha_c \frac{FU_e}{RT}\right) \right) \approx -i_0(\alpha_a + \alpha_c)\frac{FU_e}{RT},$$

$$U_e(x_1, x_2, t) \approx V_0(x_1, x_2, t) - R_c I_0 - \delta S \cdot E_3. \quad (B.3)$$

Here, $F$ is Faraday's constant, $R$ is the universal gas constant, $T$ is the absolute temperature, $\alpha_a$ is anodic empirical constant, $\alpha_c$ is cathodic empirical constant, $R_c$ is the total contact ohmic resistance, $I_0 = \delta C_c F \cdot S$ is the macroscopic electron current ($S$ is the battery cross-section area), $i_0 \approx F \cdot \chi \cdot (C_S - C_c^0)^{\alpha_a} (C_c^0)^{\alpha_c}$ is the prefactor ($\chi$ is the reaction rate constant, $C_S$ is the solubility limit of lithium in the electrode).

The following diffusion equations provide an approximate description of the electrochemical behavior of the battery cells [49]:

$$\begin{cases} \dfrac{\partial}{\partial t}\delta C_a(\mathbf{x},t) = \mathrm{div}\big(D_a\cdot\mathrm{grad}\,\delta C_a(\mathbf{x},t) - \mu_a\delta C_a\cdot\mathrm{grad}V_e(\mathbf{x},t) - C_c\mathrm{grad}(\beta_{ij}\sigma_{ij})\big), \\ \dfrac{\partial}{\partial t}\delta C_c(\mathbf{x},t) = \mathrm{div}\big(D_c\cdot\mathrm{grad}\,\delta C_c(\mathbf{x},t) + \mu_c\delta C_c\cdot\mathrm{grad}V_e(\mathbf{x},t) - C_a\mathrm{grad}(\beta_{ij}\sigma_{ij})\big). \end{cases}$$

(B.4)

The boundary conditions for ionic fluxes can be analyzed as following. When the current density at the tip electrode $x_3 = 0$ and remote bottom electrode $x_3 = h$ appeared only due to the cations ($J(\mathbf{x},t) = J_c$, $J_a = 0$), modified boundary conditions have the form:

$$-D_c\frac{\partial \delta C_c}{\partial x_3} - \mu_c\delta C_c\frac{\partial V_e}{\partial x_3} = -\frac{J_c}{eZ_c}\bigg|_{x_3=0,h},$$

$$-D_a\frac{\partial \delta C_a}{\partial x_3} + \mu_a\delta C_a\frac{\partial V_e}{\partial x_3}\bigg|_{x_3=0,h} = 0.$$

(B.5)

In decoupling approximation the concentrations are calculated ignoring the strain effects, i.e. the terms like $C_{c,a}\mathrm{grad}(\beta_{ij}\sigma_{ij})$ in Eq.(B.4), since their contribution lead into (4) is proportional to $\beta_{ij}^2$.

In the case of electrolyte electroneutrality the condition $\delta C \approx Z_a\delta C_a \approx Z_c\delta C_c$ is valid for the concentration excess ($Z_{a,c}$ are their relative charges). Then potential $V_e$ could be eliminated from Eqs.(B.4) and ambipolar diffusion equation acquires the form:

$$\frac{\partial}{\partial t}\delta C(\mathbf{x},t) = D\left(\frac{\partial^2}{\partial x_1^2} + \frac{\partial^2}{\partial x_2^2} + \frac{\partial^2}{\partial x_3^2}\right)\delta C(\mathbf{x},t).$$

(B.5)

Diffusion coefficient $D = \dfrac{D_a\mu_c + D_c\mu_a}{\mu_c + \mu_a}$. Note, that electrostatic potential does not contribute to the equation in the case of ambipolar diffusion.

Eliminating potential $V_e$ from Eqs.(B.5) and putting $h \gg R_d$ we obtained that $\dfrac{\partial}{\partial x_3}\delta C(x_1,x_2,0,t) = \dfrac{-J_c(x_1,x_2,t)}{eD(1+\mu_c/\mu_a)}$, where $J_c$ is spatially distributed renormalized ionic flux at $x_3=0$. Using expressions (B.3) for the flux $J_c$ and overpotential $U_e(x_1,x_2,t)$, the boundary conditions for the time-dependent part $\delta C(\mathbf{x},t)$ acquire the form:

$$\lambda\frac{\partial}{\partial x_3}\delta C(x_1,x_2,0,t) - \eta\delta C(x_1,x_2,0,t) = -V_0(x_1,x_2,t),$$

$$\delta C(x_1,x_2,x_3\to\infty,t)\to 0, \quad \delta C(\mathbf{x},0) = 0, \quad C_c(\mathbf{x},0) = C_c^0.$$

(B.6)

Here $\lambda$ and $\eta$ are phenomenological exchange coefficients, which can be expressed in terms of the materials constants from Eqs.(B.3), in particular $\eta = FR_c S/Z_c$ originated from the current $I_0(t) \sim \delta C_c(t)$ in the overpotential $U_e(x_1, x_2, t)$, while $\lambda = -eD\left(1+\dfrac{\mu_c}{\mu_a}\right)\dfrac{RT(C_S - C_c^0)^{-\alpha_a}(C_c^0)^{-\alpha_c}}{F^2 \chi(\alpha_a + \alpha_c)}$ is determined by the reaction rate $\chi$, and the solubility limit of lithium in the tip electrode $C_S$.

## Appendix C

For the case of the potential spatial distribution $V_0(x_1, x_2, t)$ is kept constant inside the circle of radius $R_0$ and zero outside, mechanical response temporal relaxation can be derived from Eqs.(6) and (13) as

$$u_3(0,t) = -2(1+\nu)\beta \int_0^\infty dk \, \frac{1}{2i\pi^2} \int_{A-i\infty}^{A+i\infty} ds \, \frac{R_0 J_1(kR_0) \cdot \tilde{V}_0(s) \exp(+st)}{\left(\lambda\sqrt{k^2 + s/D} + \eta\right)\left(k + \sqrt{k^2 + s/D}\right)}. \qquad (C.1)$$

The analytical results were derived for the case $\lambda=0$ and $\eta=0$ for a rectangle-like temporal dependence of the voltage pulse, $V_0(t) = V_a(\theta(t) - \theta(t-t_0))$, with pulse duration $t_0$.

For the case $\lambda=0$:

$$u_3(0,t) = -2(1+\nu)\beta \int_0^\infty dk \, R_0 J_1(kR_0) \int_0^t dt' \cdot V_0(t-t') \frac{D}{\eta}\left(\frac{\exp(-Dk^2 t')}{\sqrt{\pi D t'}} - k\cdot\left(1 - \mathrm{erf}\left(k\sqrt{Dt'}\right)\right)\right) =$$

$$= -2(1+\nu)\beta \int_0^t dt' \cdot V_0(t-t') \frac{D}{R_0 \eta}\left(\frac{R_0}{\sqrt{\pi D t'}} - \mathrm{erf}\left(\frac{R_0}{2\sqrt{Dt'}}\right)\right) = (1+\nu)\frac{\beta V_a}{\eta} \times$$

$$\begin{cases} \dfrac{1}{2\sqrt{\pi}}\left(\dfrac{4D}{R_0}\sqrt{\pi t}\,\mathrm{erf}\left(\dfrac{R_0}{2\sqrt{Dt}}\right) + \left(-8\sqrt{Dt} + R_0 \Gamma\left(-\dfrac{1}{2}, \dfrac{R_0^2}{4Dt}\right)\right)\right) \approx -\dfrac{4}{\sqrt{\pi}}\sqrt{Dt}, & t < t_0, \\[1em] \dfrac{2D}{R_0}\left(\begin{array}{l}\dfrac{R_0\sqrt{t}}{\sqrt{\pi D}}\exp\left(\dfrac{R_0^2}{4Dt}\right) + \left(\dfrac{R_0^2}{2D} + t\right)\mathrm{erf}\left(\dfrac{R_0}{2\sqrt{Dt}}\right) - \dfrac{2R_0(\sqrt{t} - \sqrt{t-t_0})}{\sqrt{\pi D}} \\ -\dfrac{R_0\sqrt{t-t_0}}{\sqrt{\pi D}}\exp\left(\dfrac{R_0^2}{4D(t-t_0)}\right) - \left(\dfrac{R_0^2}{2D} + t - t_0\right)\mathrm{erf}\left(\dfrac{R_0}{2\sqrt{D(t-t_0)}}\right)\end{array}\right) \to -\dfrac{R_0^2 t_0}{6\sqrt{\pi D}\, t^{3/2}}, & t > t_0, \end{cases}$$

(C.2)

Where the integral $\mathrm{erf}(z) = \dfrac{2}{\sqrt{\pi}} \int_0^z \exp(-x^2)\,dx$ is the Gaussian distribution, and

$\Gamma(a,z) = \int_z^\infty \exp(-x) x^{a-1} dx$ is the incomplete gamma function.

For the case $\eta=0$:

$$u_3(0,t) = -2(1+\nu)\beta \int_0^\infty dk\, R_0\, J_1(kR_0) \int_0^t dt' \cdot V_0(t-t') \frac{D}{\lambda}\left(1 - \mathrm{erf}(k\sqrt{Dt'})\right) =$$

$$-2(1+\nu)\beta \frac{D}{\lambda} \int_0^t dt' \cdot V_0(t-t') \left(1 - \exp\left(-\frac{R_0^2}{8Dt'}\right) I_1\left(\frac{R_0^2}{8Dt'}\right)\right) \approx -(1+\nu)\beta V_a \frac{R_0^2}{4\lambda} \begin{cases} \ln\left(1 + \dfrac{8Dt}{R_0^2}\right), & t < t_0; \\ \ln\left(\dfrac{R_0^2 + 8Dt}{R_0^2 + 8D(t-t_0)}\right), & t > t_0. \end{cases}$$

(C.3)

**Table D.** Lattice constants as a function of *x* and derivatives on composition x were taken from Refs.[56, 57]

| Composition | $a$ (Å)* | $c$ (Å) |
|---|---|---|
| $Li_{0.6}CoO_2$ | 2.809 | 14.32 |
| $Li_{0.7}CoO_2$ | 2.81 | 14.22 |
| $Li_{0.8}CoO_2$ | 2.809 | 14.18 |
| $Li_{0.9}CoO_2$ | 2.81 | 14.18 |
| $LiCoO_2$ | 2.81-2.82 | 14.05-14.07 |
| $LiMn_2O_4$ (spinel) | 8.24 | 8.24 |
| $LiC_6$ (Li-graphite) | 2.46 | 6.71 |

# References

[1] *Basic Research Needs for Electrical Energy Storage*, DOE BES Workshop, http://www.er.doe.gov/bes/reports/files/EES_rpt.pdf

[2] G. A. Nazri and G. Pistoia, *Lithium Batteries: Science and Technology*, Springer-Verlag, New York (2009).

[3] K. Ozawa, Editor, *Lithium-ion rechargeable batteries*, Wiley-VCH Verlag GmbH & Co. KgaA, Weinheim (2009).

[4] A. J. Bard, F.-R. F. Fan, J. Kwak, and O. Lev, *Anal. Chem.*, **61**, 132-138 (1989).

[5] T.J. Smith and K.J. Stephenson, *Electrochemical SPM: Fundamentals and Applications*, in S.V. Kalinin and A. Gruverman (Eds), *Scanning Probe Microscopy: Electrical and Electromechanical Phenomena at the Nanoscale* (v. 1&2), Springer 2006.

[6] M. Böcker, B. Anczykowski, J. Wegener, and T. E. Schäffer, *Nanotechnology*, **18**, 145505-6 (2007).

[7] Y. E. Korchev, M. Raval, M. J. Lab, J. Gorelik, C. R. Edwards, T. Rayment, and D. Klenerman, *Biophys. J.*, **78**, 2675-2679 (2000).

[8] Y. S. Cohen and D. Aurbach, *Electrochem. Commun.*, **6**, 536-542 (2004).

[9] T. Doi, M. Inabab, H. Tsuchiyac, S.-K. Jeongd, Y. Iriyamac, T. Abec, and Z. Ogumic, *J. Power Sources*, **180**, 539-545 (2008).

[10] L. Y. Beaulieu, T. D. Hatchard, A. Bonakdarpour, M. D. Fleischauer, and J. R. Dahn, *J. Electrochem. Soc.*, **150**, A1457-A1464 (2003).

[11] R. B. Lewis, A. Timmons, R. E. Mar, and J. R. Dahn, *J. Electrochem. Soc.*, **154**, A213-A216 (2007).

[12] Y. Tian, A. Timmons, and J. R. Dahn, *J. Electrochem. Soc.*, **156**, A187-A191 (2009).

[13] L. Y. Beaulieu, V. K. Cumyn, K. W. Eberman, L. J. Kraus, and J. R. Dahn, *Rev. Sci. Instr.*, **72**, 3313-3319 (2001).

[14] M. Matsui, K. Dokko, and K. Kanamura, *J. Power Sources*, **177**, 184-193 (2008).

[15] A. Clemencon, A. T. Appapillai, S. Kumar, and Y. Shao-Horn, *Electrochem. Acta*, **52**, 4572-4580 (2007).

[16] A. E. Semenov, I. N. Borodina, and S. H. Garofalini, *J. Electrochem. Soc.*, **148**, A1239-A1246 (2001).

[17] K. Kuriyama, A. Onoue, Y. Yuasa, and K. Kushida, *Surf. Sci.*, **601**, 2256-2259 (2007).

[18] R. Shao, S. V. Kalinin, and D. A. Bonnell, *Appl. Phys. Lett.*, **82**, 1869-1871 (2003).

[19] R. O'Hayre, M. Lee, and F, B. Prinz, *J. Appl. Phys.*, **95**, 8382-8392 (2004).

[20] R. O'Hayre, G. Feng, W. D. Nix, and F. B. Prinz, *J. Appl. Phys.*, **96**, 3540-3549 (2004).

[21] S.V. Kalinin, N. Balke, N.J. Dudney, and S. Jesse, Li-*ion microscopy*, patent disclosure submitted

[22] N. Balke, S. Jesse, A. N. Morozovska, E. Eliseev, D. W. Chung, Y. Kim, L. Adamczyk, R. E. García, N. Dudney, and S.V. Kalinin, *submitted to Nat. Nanotech*.

[23] N. Balke, S. Jesse, Y. Kim, L. Adamczyk, I. N. Ivanov, N. J. Dudney, and S.V. Kalinin, *submitted to Nature*.

[24] A. Gruverman and A. Kholkin, *Rep. Prog. Phys.*, **69**, 2443-2474 (2006).

[25] A. Gruverman and S. V. Kalinin, *J. Mat. Sci.*, **41**, 107-116 (2006).

[26] . V. Kalinin, B. J. Rodriguez, S. Jesse, B. Mirman, E. Karapetian, E. A. Eliseev, and A. N. Morozovska, *Annu. Rev. Mat. Sci.*, **37**, 189-238 (2007).

[27] A. Gruverman, O. Auciello, and H. Tokumoto, *J. Vac. Sci. Technol. B*, **14**, 602-605 (1996).

[28] J. R. I. Lee, T. Y.-J. Han, T. M. Willey, D. Wang, R. W. Meulenberg, J. Nilsson, P. M. Dove, L. J. Terminello, T. van Buuren, and J. J. De Yoreo, J. *Amer. Chem. Soc.*, 129, 10370-10381 (2007).

[29] B. J. Rodriguez, S. Jesse, and S. V. Kalinin, *Phys. Rev. Lett.*, **96**, 237602 (2006).

[30] B. J. Rodriguez, S. Jesse, A. P. Baddorf, S. H. Kim, and S. V. Kalinin, *Phys. Rev. Lett.*, **98**, 247603 (2007).

[31] F. Felten, G. A. Schneider, J. Munoz Saldana, and S. V. Kalinin, *J. Appl. Phys.*, **96**, 563-568 (2004).

[32] D. A. Scrymgeour and V. Gopalan, *Phys. Rev. B,* 72, 024103 (2005).

[33] E. A. Eliseev, S. V. Kalinin, S. Jesse, S. L. Bravina, and A. N. Morozovska, *J. Appl. Phys.*, **102**, 014109 (2007).

[34] A. N. Morozovska, E. A. Eliseev, S. L. Bravina, and S. V. Kalinin, *Phys. Rev. B*, **75**, 174109 (2007).

[35] P. M. Gomadam, J. W. Weidner, R. A. Dougal, and R. E. White, *J. Power Sources*, **110,** 267-284 (2002).

[36] G. G. Botte, V. R. Subramanian, and R. E. White, *Electrochimica Acta*, **45**, 2595-2609 (2000).

[37] D. Zhang, B. N. Popov, and R. E. White, *Journal of The Electrochemical Society*, **147**, 831-838 (2000).

[38] X. Zhang, W. Shyy, and A. M. Sastrya, *Journal of The Electrochemical Society*, **154**, A910-A916 (2007).

[39] X. Zhang, A. M. Sastrya, and W. Shyy, *Journal of The Electrochemical Society,* **155**, A542-A552 (2008).

[40] S. Suresh, *Fatigue of Materials*, Cambridge University Press, Cambridge (1998).

[41] J. F. Nye, *Physical Properties of Crystals*, Oxford University Press, Oxford (1998).

[42] S. P. Timoshenko and J. N. Goodier, *Theory of Elasticity*, McGraw-Hill, New-York (1970).

[43] L. D. Landau and E. M. Lifshitz, *Theory of Elasticity. Theoretical Physics,* Vol. 7, Butterworth-Heinemann, Oxford, (1976).

[44] T. Mura, *Micromechanics of defects in solids*, Martinus Nijhoff Publishers, Boston (1987).

[45]. J. S. Newman, *Electrochemical Systems*, Prentice Hall International, Englewood Cliffs, New Jersey (1980).

[46] R. E. García, Y.-M. Chiang, W. C. Carter, P. Limthongkul, and C. M. Bishop, *J. Electrochem. Soc.*, **152**, A255-A263 (2005).

[47]. R. E. García, C. M. Bishop, and W. C. Carter, *Acta Materialia*, **52**, 11–21 (2004).

[48]. W. C. Carter, J. E. Taylor, and J. W. Cahn, *Journal of Materials*, **49**, 30-36 (1997).

[49] C. Brissot, M. Rosso, J. N. Chazalviel, and S. Lascaud, *J. Electrochem. Soc.,* **146,** 4393-4400 (1999).

[50] H. S. Carslaw and J. C. Jaeger, *Conduction of heat in solids*, Clarendon Press, Oxford (1959).

[51] B. J. Rodriguez, S. Jesse, K. Seal, A. P. Baddorf, S. V. Kalinin, and P. D. Rack, *Appl. Phys. Lett.*, **91**, 093130 (2007).

[52] P. L. T. M. Frederix, M. R. Gullo, T. Akiyama, A. Tonin, N. F. de Rooij, U. Staufer, and A. Engel, *Nanotechnology*, **16**, 997 (2005).

[53] C. Kranz, G. Friedbacher, B. Mizaikoff, A. Lugstein, J. Smoliner, and E. Bertagnolli, *Anal. Chem.*, **73**, 2491 (2001).

[54] J. V. Macpherson and P. R. Unwin, *Anal. Chem.*, **72**, 276 (2000).

[55] J. Xie, N. Imanishi, T. Matsumura, A. Hirano, Y. Takeda, and O. Yamamoto, *Solid State Ionics*, **179**, 362-370 (2008).

[56] J. Cho, Y. J. Kim, and B. Park, *Chem. Mater.*, **12**, 3788-3791 (2000).

[57] J. Cho, Y. J. Kim, T.-J. Kim, and B. Park, *Angew. Chem. Int. Ed.* **40**, 3367 (2001).

[58] T. Ohzuku and Y. Makimura. *Res. Chem. Intermed.*, **32**, 507–521 (2006).

[59] P. Maksymovych, S. Jesse, P. Yu, R. Ramesh, A. P. Baddorf, and S. V. Kalinin, *Science*, **324**, 1421-1425 (2009).

[60] S. Jesse, A. P. Baddorf, and S. V. Kalinin, *Nanotechnology*, **17**, 1615-1628 (2006).

[61] C. Harnagea, M. Alexe, D. Hesse, and A. Pignolet, *Appl. Phys. Lett.*, 83, 338-340 (2003).

[62] A. Salehi-Khojin, S. Bashash, N. Jalili, G. L. Thompson, and A. Vertegel, *J. Dyn. Syst.-T. ASME*, **131**, 061107 (2009).

[63] S. V. Kalinin and D. A. Bonnell, *Phys. Rev. B*, **65**, 125408 (2002).

[64] U. Rabe, V. Sherer, S. Hirsekorn, and W. Arnold, *J. Vac. Sci. Technol. B*, **15**, 1506 (1997).

[65] J. A. Turner, S. Hirsekorn, U. Rabe, and W. Arnold, *J. Appl. Phys.*, **82**, 966 (1997).

[66] D. Sarid, *Scanning Force Microscopy: With Applications to Electric, Magnetic, and Atomic Forces*, Oxford University Press, Oxford (1994).

[67] B. Mirman and S.V. Kalinin, *Appl. Phys. Lett.*, **92**, 083102 (2008)

[68] S. Jesse, B. Mirman, and S. V. Kalinin, *Appl. Phys. Lett.*, **89**, 022906 (2006).

[69] R. Proksch and S.V. Kalinin, *Piezoresponse Force Microscopy*, Asylum Research support note, http://www.asylumresearch.com/Applications/PFMAppNote/PFMAppNote.shtml

[70] B. J. Rodriguez, C. Callahan, S. V. Kalinin, and R. Proksch, *Nanotechnology*, **18**, 475504 (2007).

[71] S. Jesse, S. V. Kalinin, R. Proksch, A. P. Baddorf, and B. J. Rodriguez, *Nanotechnology*, **18**, 435503 (2007)

[72] A. J. Bard and L. R. Faulkner, *Electrochemical Methods: Fundamentals and Applications*, John Wiley & Sons, New York (2001).

[73] S. V. Kalinin, B. J. Rodriguez, S. Jesse, K. Seal, R. Proksch, S. Hohlbauch, I. Revenko, G. L. Thompson, and A. A. Vertegel, *Nanotechnology*, **18**, 424020 (2007).

[74] G. M. Sacha, A. Verdaguer, and M. Salmeron, *J. Phys. Chem. B*, **110**, 14870-14873 (2006).